\documentclass[pdftex,twocolumn,10pt,letterpaper]{extarticle}

\pdfoutput=1 

\newcommand{\AUTHORS}{Osama Haq\textsuperscript{1}, Cody Doucette\textsuperscript{2}, John W. Byers\textsuperscript{2}, and Fahad R. Dogar\textsuperscript{1}}
\newcommand{\TITLE}{CASPR: Judiciously Using the Cloud for \\Wide-Area Packet Recovery}

\newcommand{\KEYWORDS}{Put your keywords here}
\newcommand{\CONFERENCE}{Somewhere}
\newcommand{\PAGENUMBERS}{yes}       
\newcommand{\COLOR}{yes}
\newcommand{\showComments}{no}
\newcommand{\comment}[1]{}
\newcommand{\onlyAbstract}{no}

\newfont{\ttlfnt}{phvb8t at 18pt}  
\newfont{\aufnt}{phvr8t at 12pt}    
\newfont{\auit}{phvro8t at 12pt}    
\newfont{\affaddr}{phvr8t at 10pt}  

\usepackage[T1]{fontenc}
\usepackage[utf8]{inputenc}
\usepackage{verbatim}
\usepackage{newtxtext,newtxmath}       
\usepackage{bm}                        
\usepackage[scaled=0.92]{helvet}
\usepackage{courier}
\usepackage{subfigure}
\usepackage[ruled,noline,noend]{algorithm2e}
\usepackage{supertabular}

\usepackage{wrapfig}

\usepackage{ifthen}
\ifthenelse{\equal{\PAGENUMBERS}{yes}}{%
\usepackage[nohead,
            left=1.0in,right=1.0in,top=1.0in,
            footskip=0.5in,bottom=1in,     
            columnsep=0.33in
		]{geometry}
}{%
\usepackage[noheadfoot,left=0.75in,right=0.85in,top=0.75in,
            footskip=0.5in,bottom=1in,
            columnsep=0.25in
	    ]{geometry}
}

\usepackage[font=bf]{caption}

\usepackage{titlesec}
\titlespacing{\paragraph}{0pt}{*1}{*1}      
\titleformat{\paragraph}[runin]{\normalfont\normalsize\bfseries}{\theparagraph}{1em}{}

\titleformat{\section}
  {\bf\large\uppercase}{\thesection.\quad}{0pt}{}
\titleformat{\subsubsection}
  {\it\large}{\thesubsubsection\quad}{0pt}{}

\usepackage{enumitem}
\setlist{itemsep=0pt,parsep=0pt}             

\usepackage{fancyhdr}

\ifthenelse{\equal{\PAGENUMBERS}{yes}}{%
  \pagestyle{plain}
}{%
  \pagestyle{empty}
}

\usepackage[numbers]{natbib}

\usepackage{booktabs}
\usepackage{multirow}
\usepackage{color}
\usepackage{colortbl}
\usepackage{float}                           

\ifthenelse{\equal{\COLOR}{yes}}{%
  \usepackage[colorlinks]{hyperref}
}{%
  \usepackage[pdfborder={0 0 0}]{hyperref}
}
\usepackage{url}

\hypersetup{%
pdfauthor = {\AUTHORS},
pdftitle = {\TITLE},
pdfsubject = {\CONFERENCE},
pdfkeywords = {\KEYWORDS},
bookmarksopen = {true}
}

\usepackage{hyperref}
\hypersetup{
     colorlinks   = true,
     citecolor    = blue,
     urlcolor = blue,
}

\definecolor{placeholderbg}{rgb}{0.85,0.85,0.85}

\usepackage[pdftex]{graphicx} 
\usepackage{soul} 
\soulregister\cite7
\soulregister\ref7
\soulregister\pageref7

\usepackage[absolute]{textpos}

\newfloat{acmcr}{b}{acmcr}

\newcommand{\note}[2]{
    \ifthenelse{\equal{\showComments}{yes}}{\textcolor{#1}{#2}}{}
}

\newcommand{\fahad}[1]{\note{blue}{Fahad: #1}}
\newcommand{\john}[1]{\note{red}{John: #1}}
\newcommand{\osama}[1]{\note{blue}{Osama: #1}}
\newcommand{\cody}[1]{\note{blue}{Cody: #1}}
\newcommand{\sys}{CASPR} 

\date{}
\title{\ttlfnt \TITLE }
\author{{\aufnt \AUTHORS \vspace{1ex}}\\
{\affaddr \textsuperscript{1}Tufts University, \textsuperscript{2}Boston University}}

\begin{document}

\maketitle


\begin{abstract}
 

We revisit a classic networking problem -- how to recover from lost packets in the best-effort Internet. 
We propose \sys{}, a system that \emph{judiciously} leverages the cloud to recover from lost or delayed packets. 
\sys{} supplements and protects best-effort connections by sending a small number of coded packets along the highly reliable but expensive cloud paths. 
When receivers detect packet loss, they recover packets with the help of the nearby data center, not the sender, thus providing quick and reliable packet recovery for latency-sensitive applications.  
Using a prototype implementation and its deployment on the public cloud and the PlanetLab testbed, 
we quantify the benefits of \sys{} in providing fast, cost effective packet recovery. 
Using controlled experiments, we also explore how these benefits translate into improvements up and down the network stack.

\end{abstract}

\ifthenelse{\equal{\onlyAbstract}{no}}{%
\section{Introduction}
\label{sec:intro}

Notable in an era of constant change, the Internet of today retains a 40-year old design philosophy built around the principle of a ``stupid network'' with smart endpoints.  Today's IP network layer offers best effort service, with no guarantees on latency, packet loss, or bandwidth; while important functions like reliability and congestion control continue to be implemented at the end-points, as part of the TCP transport.  This is despite well known pain points in many scenarios -- for example, recent studies again highlight how network problems lead to problems for latency sensitive applications, such as streaming~\cite{qoe1, qoe2} and short web-transfers~\cite{gentleaggression}.
With the emergence of virtual reality (VR) and augmented reality (AR) applications, the demands on the network continue to increase and apply pressure to the underlying architecture.




Fortunately, the emergence of \emph{the cloud} offers us a new opportunity to better support the needs of latency sensitive applications without re-architecting the underlying best effort Internet. By the cloud, we refer to a distributed network of data centers (DCs), inter-connected through a private network (e.g., Azure, EC2, Google Cloud).  For any communication between two end-points, we can potentially use the cloud as an \emph{overlay}, with DCs acting as an insertion point for in-network services~\cite{rewanhotnets, via2016}. A cloud-based overlay offers unique opportunities: cloud paths are well-provisioned, offering low jitter and very high reliability; and each DC has visibility into many users and applications, so it can act as a unique vantage point for control. On the flip side, using the cloud as an overlay can be costly: cloud providers charge for the use of their resources (e.g., processing, network connectivity), with wide area network (WAN) bandwidth being particularly expensive~\cite{swan, eurosys15, rewanhotnets}. Therefore, we argue that the most effective use of the cloud as an overlay is one that does so in a \emph{judicious} manner. 

In this paper, we design, implement, and evaluate \sys{}\footnote{Cloud ASsisted Packet Recovery (\sys{}).}, a packet recovery service for latency-sensitive applications operating on the wide-area best-effort Internet. \sys{} uses the cloud overlay \emph{only} for packet recovery: it sends a low rate of \emph{coded} packets across the inter-DC paths; these coded packets are used only in case of a packet loss on data transmitted along the best effort Internet path. 
The coded packets provide resilience to a wide range of common packet loss and jitter scenarios, e.g., both random and bursty losses, and packets experiencing high delay. At the same time, it makes economical use of the expensive inter-DC bandwidth. The two key pieces of \sys{}'s design are the \emph{coding scheme}  and the receiver-driven \emph{cooperative recovery protocol}, described next.

Using \sys{}, each packet sent on the public Internet path is additionally routed to the source's nearest ingress DC.  Each ingress DC
then determines a coding plan on-the-fly based on the arrival of all packets to that DC.  
Packets destined to the same destination region -- with a common {\em egress DC} --  that arrive within a short window of time
are grouped together to form
a batch, over which coding occurs. For each batch of $b$ packets, \sys{} generates a set of encoding packets at a tunably low rate $r < 1$, so that $rb < b$ packets are computed and transmitted along the DC-DC path and protect each batch of $b$ best-effort transmissions.  Encoding is done in two complementary forms:  cross-stream coding encodes packets \emph{across} user streams, while in-stream coding encodes packets \emph{within} a single flow.  The latter provides protection against random losses, while the former can protect against burst losses or even a sustained outage along a best-effort path, as we demonstrate shortly.


\sys{} uses a receiver-driven mechanism to initiate and cooperatively recover from packet loss. 
To initiate recovery, the receiver notifies the egress DC about the missing packets.  The egress DC consults the coding plan to determine the right recourse for recovery -- forwarding in-stream coding packets while sufficient and failing over to 
cooperative recovery
 when necessary. For packets requiring cooperative recovery,  the egress DC must gather the rest of the batch: the cross-stream packet(s) plus the data packets from other receivers.  
 Our cooperative recovery mechanism exploits today's cloud bandwidth pricing model, in which incoming bandwidth is typically free~\cite{amazonprice,azureprice,googleprice} -- so recovery incurs relatively low bandwidth cost as receivers send their data packets to the DC using the free ingress bandwidth. 

We have implemented \sys{} as a network layer service that intercepts transport segments and performs recovery as described above, 
seamlessly working with both TCP and UDP based applications without requiring any application modification. 
We have deployed \sys{} as a service on a public cloud for over a month and measured the wide area performance of PlanetLab paths. Our results show that \sys{} is able to recover more than 70\% of losses, the recovery is typically within half a round-trip time (RTT),
and the overhead of using the cloud judiciously is far less than a traditional overlay solution that uses the cloud exclusively.
Through controlled experiments, we also evaluate how \sys{} interacts with protocols up and down the network stack -- we show that: i) \sys{}'s packet recovery can improve the user's QoE experience for a skype video conferencing scenario, ii) \sys{} can speed up short web transfers by avoiding TCP timeouts and congestion avoidance caused by bursty losses, iii) it is feasible to use \sys{} on mobile networks, in terms of bandwidth, energy consumption, and latencies to nearby DCs.





Overall, we make the following contributions in this paper. 

\begin{itemize}[leftmargin=*]
    \item A case for judicious packet recovery by sending only coded packets across the inter-DC path. 
    
    \item A design and implementation of \sys{}, including a practical, tunable coding module and a receiver-driven protocol for cooperative packet recovery. 
    
    \item A multi-faceted evaluation of \sys{}, using both network and user level metrics, on diverse networks (PlanetLab, cellular) and applications (video conferencing, short web transfers).
    

\end{itemize}


\section{Motivation}
\label{sec:motivation}


\subsection{Problem Description}
\label{subsec:prob-desc}
Our focus is on latency-sensitive applications, such as video chat, collaborative AR/VR, and short web transfers. Unfortunately, network impairments, such as packet loss and jitter,
 still remain a problem for these applications. Recent studies show how poor network conditions impact the user experience for popular applications. 
For example, Microsoft's analysis of 430 million Skype calls reveals that network conditions such as round trip time (RTT), loss rate, and jitter are directly correlated with perceived user experience~\cite{via2016}. They observe that poor network conditions impact 15\% of calls, with wide area paths being 2-3x more prone to poor network conditions. Similar problems have also been observed for short web transfers~\cite{gentleaggression}. 

 
While network problems can be of different types, we focus on 
the particularly challenging scenario of impairments in the ``middle'' for wide-area paths -- paths with high round-trip time between them (e.g., across continent, east-west coast, etc).
For such wide-area paths, even if the two end-points have good access network connectivity (e.g., enterprises, universities, etc) -- as we assume in this paper --  they still have to rely on the best effort Internet (or the ``middle''), which can experience poor performance due to various issues in the core (e.g. routing misconfigurations~\cite{arrow}, peering disputes~\cite{congestion-imc14}, router/link failure~\cite{ron,arrow}, etc).

\paragraph{Limitations of Existing Solutions.}Several prior proposals, from the seminal work on RON~\cite{ron}, to recent proposals like Arrow~\cite{arrow}, also focus on this problem of network impairments in the ``middle''.
However, solving this problem is challenging. 
Simple retransmission based techniques (e.g., TCP), as well as edge proxy solutions~\cite{tapaconext, Bakre97, catnap, DogarM208}, have limited benefits given our target application and network settings (latency sensitive applications, wide area paths, losses in the middle, etc). 

Redundancy-based solutions (e.g., FEC) 
are more applicable, but unfortunately cannot cope with burst losses or outages that can occur on wide area Internet paths. Previous measurement studies as well as our experiments show that such bursty losses and outages can last for seconds~\cite{ron, rewanhotnets}. Other proposals (e.g. Maelstrom~\cite{maelstrom}, Raptor~\cite{raptor}) use layered interleaved coding to provide protection against bursty losses, but their reliance on future packets adds non-tolerable delay for latency sensitive applications. 

Finally, overlay networks, such as RON~\cite{ron}, can potentially avoid these bursty losses and outages on the direct Internet path by taking alternate (overlay) paths. However, traditional overlay networks have to deal with problems of node churn and lack of performance predictability, which limit their practical use for performance critical applications.

\subsection{The Cloud as an Overlay}
\label{subsec:cloud-overlay}
We consider using a cloud overlay as a potential solution to our problem. Some interactive applications, such as Skype and Google Hangout, are already migrating their services to at least partial use of cloud relays~\cite{via2016}, but there has been little work in studying how to best utilize the cloud for these types of interactive communications. Therefore, we characterize the properties of cloud paths in terms of network conditions and cost, and ask:
can the cloud be leveraged in a cost-efficient way to make up for the Internet's performance limitations?

\emph{Benefits.} There are several advantages of using the cloud as an overlay. First, measurements show that cloud paths are highly reliable with a typical downtime target of a few minutes per month~\cite{goog-avail, cloudstudy-www}. A recent study shows that inter data center paths have an order of magnitude lower loss rate, and significantly higher bandwidth, compared to public Internet paths~\cite{cloudstudy-www}. Similar benefits are being extended all the way up to ISP networks, with major cloud operators providing bandwidth-guaranteed pipes between their data centers and customer premises (e.g., Azure ExpressRoute~\cite{azure-expressroute}, AWS Direct Connect~\cite{aws-direct-connect}).  These advances are poised to make the \emph{entire} cloud overlay highly reliable, including both the WAN as well as the last hop to the end-users.  
Second, cloud infrastructure provides the ability to implement in-network services in software in a  scalable and fault tolerant fashion, with the help of network function virtualization (NFV)~\cite{combsekar2012,middleboxsherry}. 
Third, cloud operators also strive to provide low latency access to end users, for example, by direct peering with customer ISPs to provide better and faster access to popular web services~\cite{goog-onehop}. 
Recent studies
show that using cloud paths only adds a small amount of latency compared to that of the public Internet~\cite{middleboxcloud, rewanhotnets}.

\emph{Cost.} Although the cloud as an overlay provides significant benefits, it can be expensive to use, especially due to the high cost of inter-DC bandwidth. 
Anecdotal evidence, as well as our discussions with operators, suggests that an inter-continental leased line could be an order of magnitude or more expensive compared to a connection to the best effort Internet.
This reasoning underlies several recent proposals that try to make efficient use of inter-DC bandwidth in order to reduce their network costs~\cite{rewanhotnets,interdcpricing2016,netstitcher2011}. 

\paragraph{A Cloud-based Packet Recovery Service.}
We argue that in current settings, we only need to rely on the cloud whenever the best-effort Internet cannot provide the desired reliability. 
\begin{figure}[!t]
\centering
\includegraphics[width=1.4in]{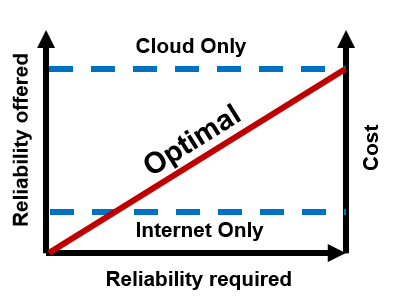}
\caption{Reliability Solution Space Analysis}
\label{fig:solution-space}
\end{figure}
We propose \emph{judicious} use of cloud paths: leveraging the availability and performance of cloud only when the best-effort Internet encounters problems. 

Figure~\ref{fig:solution-space} shows why a judicious use of the cloud is desirable.
An Internet only solution has low cost but offers low reliability. 
In contrast, a cloud only solution offers high reliability, at the expense of higher cost. 
By judiciously using the cloud, we can approach the optimal line:  applications can rely on the public Internet to deliver most of the packets, and use the cloud to augment end-to-end availability. 
For example, an application using an Internet path with 1\% loss is still getting 99\% of its packets delivered, so it should sparingly use the cloud. 
Our proposed packet recovery service is an example of how this can be done: it uses the cloud \emph{only} for packet recovery.

\section{Design}
\label{sec:design}
In this section, we first outline key design goals for a packet recovery service.
Next we overview \sys{}'s design and show how it meets these challenges, and then describe key components of our system in detail. Our goals are:


   
    \paragraph{Timely packet recovery.}  Traditional retransmission mechanisms for end-to-end packet recovery necessarily incur latency overheads at the timescale of multiple RTTs. For applications with much lower latency tolerance, recovering both delayed packets as well as packets that are lost on sub-RTT timescales is a requirement.
    
     \paragraph{Maximal protection.} Traditional methods for protecting against bit errors and packet losses have limited efficacy when faced with loss bursts or 
     outages. Any service should provide protection against a wide range of loss scenarios, minimizing the number of packets that are lost or excessively delayed. 
     
    \paragraph{Low bandwidth overhead (cost).} Minimize the use of the expensive inter-DC cloud bandwidth, thus ruling out traditional overlay solutions which use the cloud path for \emph{all} of the traffic~\cite{via2016}. 

\begin{figure}[!t]
 \includegraphics[scale=0.5]{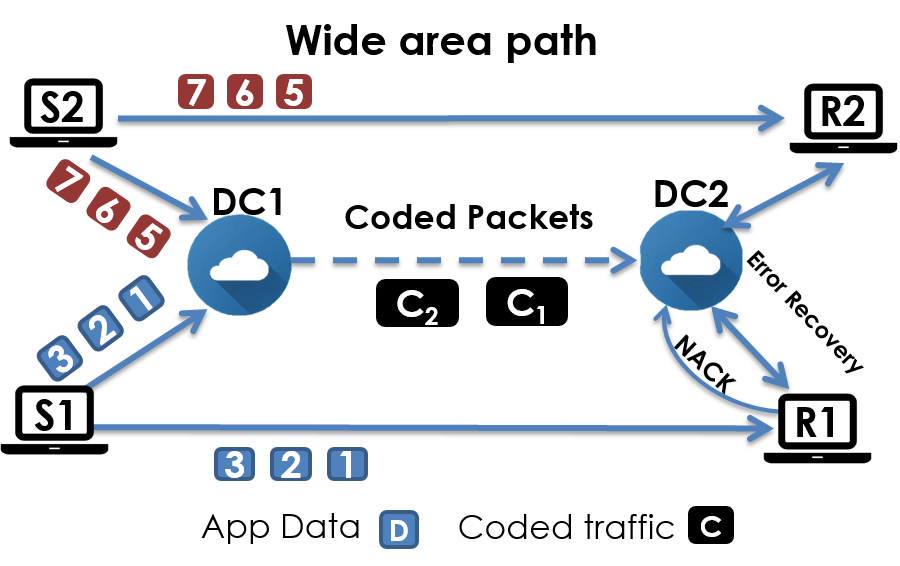}
 \caption{\sys{} Overview.}
 \label{fig:overview}
\end{figure}

\subsection{\sys{} Overview}

Figure~\ref{fig:overview} shows the high level workings of \sys{} through a basic example. There are two sender-receiver pairs ($S_1$-$R_1$ and $S_2$-$R_2$), 
each connected via a wide area path. For example, they may be on separate continents. There are two DCs that are running the \sys{} service: DC1 is in close physical proximity to the senders and DC2 is close to the receivers. 
The senders and receivers communicate on the wide area path using IP's best effort service, but in addition, each sender also sends a copy of each packet
to its nearby \sys{} service (DC1). 


\paragraph{Coding.} DC1 generates a small number of \emph{coded} packets, which are sent to the recovery service at DC2 using the inter-DC cloud path. \sys{} uses a novel \emph{cross-stream} coding design: coding is done \emph{across} a subset of user streams, which protects against bursty losses or even complete outages on a network path. For example, if ($S_1$-$R_1$) experiences an outage, \sys{} undertakes a \emph{cooperative} recovery process by combining the coded packets at DC2 with the data packets of $S_2$-$R_2$ to recover the lost packets. 

The cooperative recovery process, however, has its own set of challenges. 
First, decoding overhead can be high since that involves getting data packets from all other flows in the encoding subset. 
To ensure that this procedure is invoked only when necessary, \sys{} also uses \emph{in-stream} coding,
whereby it generates a small number of FEC packets \emph{within} a single user stream, thereby avoiding the potentially costly cooperative recovery process for random losses.
Unlike traditional end-to-end FEC, these in-stream coded packets are transmitted along the inter-DC path. Second, during cooperative recovery, some packets could be lost or delayed, especially if many streams are involved -- we call this the \emph{straggler problem}. \sys{}'s cross-stream coding accounts for potential stragglers by generating extra coded packets, thereby treating packets from stragglers similarly to losses on the direct Internet path.  

\paragraph{Packet Recovery.} \sys{} uses a novel, 
receiver-driven packet recovery protocol which optimizes for
today's cloud pricing model that only charges for egress cloud bandwidth usage. Unlike traditional protocols, there is no involvement of the sender; instead, the receiver undertakes out-of-band packet recovery; it detects losses using a combination of missing sequence numbers and timeouts, and informs DC2 about any lost packets. For some packet losses, it suffices for DC2 to relay the in-stream coded packets. In other instances, the DC may need to undertake cooperative recovery, by contacting other receivers, obtaining their packets and combining them with the cross-stream coded packets to recover missing packets. Since most of the bandwidth overhead in this process is due to packets coming \emph{into} the cloud (DC2) and is thus cost-free,
we proactively use cooperative recovery after three consecutive lost packets, resulting in faster recovery for large bursty losses or outages. 


\paragraph{Cost.} \sys{}'s design explicitly minimizes the bandwidth cost of using the cloud's egress bandwidth. This is reflected in two design decisions. First, \sys{} only exchanges a small number of recovery packets between DCs. This is the major difference between \sys{} and a solution that \emph{only} uses the cloud~\cite{via2016}, we also compare cost of both these solutions in ~\S\ref{subsec:prototype-eval}. Second, recovery between DC2 and the receiver is \emph{on-demand}: this saves the egress bandwidth charges at DC2, albeit at the expense of (slightly) slower recovery.
In general, the design of \sys{} aims to explore the spectrum between exclusive use of best-effort Internet paths and exclusive use of cloud paths, striving to optimize the trade-off between wide area packet recovery and cost.

\subsection{Coding}
\label{sec:coding}

We now elaborate the two key decisions made by the encoding process: i) deciding which batches of packets to encode over (Coding Plan) and ii) deciding how many encoded packets to generate (Coding Rate).

\subsubsection{Coding Plan}

The coding plan needs to account for \emph{spatial} and \emph{temporal} constraints while forming a batch of packets on which coding will be applied. By spatial constraints, we mean that only flows with the same destination DC can be considered together for cross-stream coding. For example, if DC1 is in the Eastern US region and is receiving traffic destined for a European DC and an Asian DC, it forms two groups, one for each destination DC.
Each flow belongs to one group and DC1 keeps a track of the mapping of flows to groups. 
Within a group, we pick a further subset of flows based on the arrival timing of their packets to form coding batches. 

Temporal constraints restrict packets in a batch to only those packets that arrive within a short interval -- this imposes an encoding delay. For in-stream coding, the encoding delay is well-understood (and is considered a limitation of FEC for low bitrate applications) as we need to wait for all packets in a block to arrive before we can generate the FEC packets. However, \sys{}'s use of  cross-stream coding ensures that encoding delay is typically lower, because packets from different user streams can arrive within a short time-frame, even if each application individually is generating low bitrate traffic. Finally, our coding module limits the block size (for a given level of protection) and uses timeouts to bound delay. 

\subsubsection{Coding Rate}
Given a batch of data packets arriving at DC1, \sys{} needs to decide how many cross-stream and in-stream coded packets to generate. For both types, the coded packets are created using a block code (for example, Reed-Solomon codes), which allows \sys{} to generate multiple coded packets per batch if desired.  Figure~\ref{fig:coding-1} depicts some of the possible tradeoffs, for a batch of 20 packets from four synchronous (for simplicity) flows, A-D.  In this depiction, in-stream encoding proceeds horizontally: a single FEC packet ($Y_i$) is produced for each flow $i$.  Cross-stream encoding proceeds vertically:  two cross-stream packets are produced from groups of four packets across flows, i.e., $A2$, $B2$, $C2$, and $D2$ are combined to generate coded packets $X3$ and $X4$.

Coding logically proceeds with two rates: an in-stream encoding rate of $s < 1$ coded packets per within-flow data packets, and a cross-stream encoding rate of 
$r < 1$ coded packets per data packet, where the data packets are selected among at most $k$ different flows.\footnote{We deviate from the standard notation of block coding theory, where $k$ data elements are encoded to generate a block of size $n$, yielding $(n - k)$ coded packets. Data rate and timing constraints may require us to code before $k$ packets are available.} 
Note that DC1 must also include information in the coded packets about which flows and sequence numbers are represented, to facilitate later recovery.
In our depicted setting, we have $k = 4$, $r = \frac{2}{4}$ and $s=\frac{1}{5}$, but in practice we use fewer coded packets for a batch of data packets, with the typical overhead of coded packets less than 20\%. 

Coded packets provide protection in multiple ways.  In-stream encoding packets protect primarily against random loss, much like traditional FEC.  As depicted in Figure~\ref{fig:coding-2}, packet $Y_A$ can recover from the loss of $A_3$.
Cross-stream packets protect against bursty losses or outages on the direct Internet path, and against the possibility of receivers being stragglers during cooperative recovery.
In figure~\ref{fig:coding-4}, if some of $C$'s packets are also lost on the direct path, additional protection using more encoding packets could enable recovery at both $A$ and $C$.  The cooperative recovery mechanism is described in \ref{subsec:loss-recovery}.


Our depiction begs the question: why are both types of encoding useful, and how much protection is advisable?   We view in-stream coding as a first line of defense: providing faster recovery for random losses.  Cross-stream encoding, on the other hand is both much more powerful (it can recover both random and bursty losses), but also incurs a potentially higher delay associated with recovery.
In our evaluation, we quantify these costs and weigh them against how beneficial multiple coded packets are in terms of protection from losses and decreased packet recovery times.

\begin{figure}[!t]
\centering
  \hspace{-2em}
  \subfigure[][\small cross-stream and 
  in-stream encoding]{\includegraphics[width=1.13in]{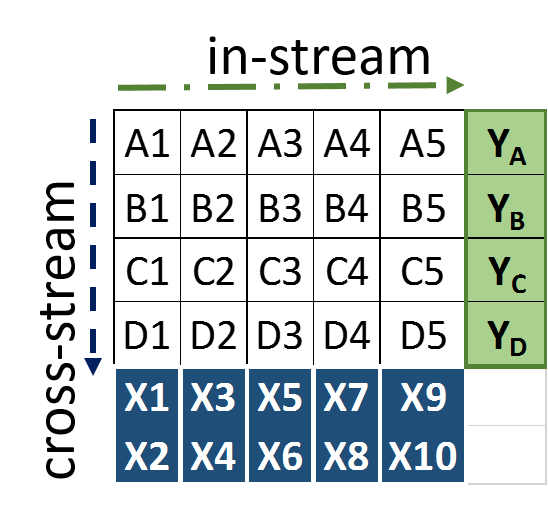}\label{fig:coding-1}}
\hspace{2em}
  \subfigure[][\small $Y_A$~protects~flow
  $A$ (in-stream).]{\includegraphics[width=1.0in]{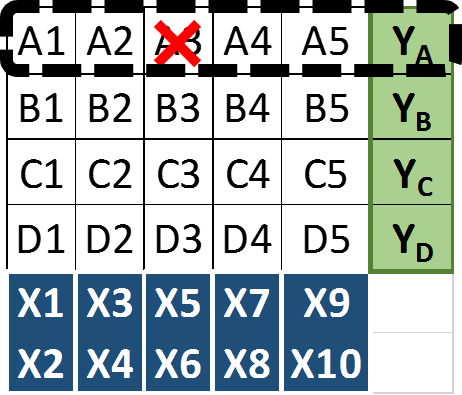}\label{fig:coding-2}}
 \hspace{0.35em}
  \subfigure[][\small $X$'s protect flow $A$ (cross-stream).]{\includegraphics[width=1.0in]{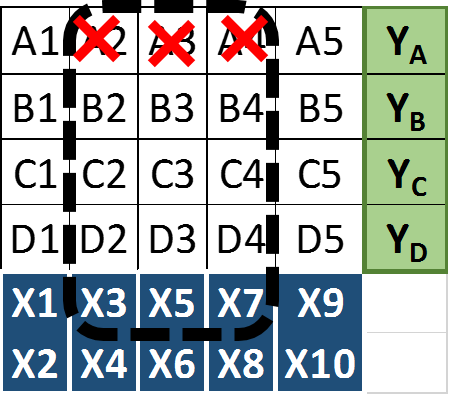}\label{fig:coding-3}}
\hspace{2em}
 \subfigure[][\small $X$'s protect $A$
 and $C$ (cross-stream).]{\includegraphics[width=1.0in]{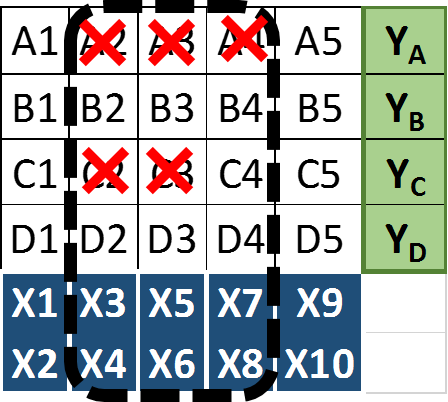}\label{fig:coding-4}}
  \caption{Coded Packet Generation and Recovery.}
 \end{figure}
\paragraph{Coding Parameters.} For cross-stream coding, we use a default of two cross-stream coded packets ($r = 2/k$) to mitigate the effects of stragglers and protect against bursty losses and outages.
In practice, we bound $k$ to a moderate value ($k <= 10$ in our evaluation), since larger values add significant overhead in the cooperative recovery process. When more than $k$ flows use \sys{} concurrently at an ingress DC, the DC organizes them into subgroups of at most $k$ flows per group.

For in-stream coding, we find that for interactive applications -- where the average frame rate is 10-15 fps and the average frame is composed of 2-5 packets~\cite{chitchat} -- it is best suited to send an in-stream packet for each frame ($s=\frac{1}{5}$), although that results in relatively higher overhead, so applications with low cost budget can choose to fall back to cross-stream coding only. The in-stream encoding overhead is less for applications that send back-to-back packets, such as 
TCP flows, where a single coded packet can be sent for an entire TCP window (e.g., $s=\frac{1}{16}$ or $s=\frac{1}{32}$)




\begin{algorithm}[!t]
\scriptsize
\SetKwFunction{KwFn}{dc1\_process(pkt, flow\_id):}
\textbf{def} \texttt{in\_stream\_qs[]}\\
\textbf{def} \texttt{cross\_stream\_qs[][]}\\
\BlankLine
\KwFn

\Indp
   \BlankLine
   \textit{// (1) In-stream coding.}\\
   \nl \texttt{q = in\_stream\_qs[flow\_id]}\\
   \nl \texttt{q.push(pkt)}\\
   \nl \If{\texttt{q.isFull()}} {
       \nl \texttt{in\_coded\_pkts = encode(q)}\\
       \nl \texttt{send(dc2\_id, in\_coded\_pkts)}\\
    }
    \BlankLine
    \textit{// (2) Cross-stream coding.}\\
    \nl \texttt{dc2\_id = extract\_dc2\_id(flow\_id)}\\
    \nl \texttt{q\_index = next\_round\_robin\_q(flow\_id)}\\
    \nl \texttt{q = cross\_stream\_qs[dc2\_id][q\_index]}\\
    \BlankLine
    \textit{// Find a queue that doesn't have a packet from this flow.}\\
    \nl \texttt{initial\_q = q}\\
    \nl \While {\texttt{q.contains(flow\_id)}} {
        \nl \texttt{q\_index = next\_round\_robin\_q(flow\_id)}\\
        \nl \texttt{q = cross\_stream\_qs[dc2\_id][q\_index]}\\
        \textit{// If we've tried all q's, empty the first by encoding or discarding.}\\
        \nl \If {\texttt{q $==$ initial\_q}} {
            \nl \If {\texttt{q.size() > 1}} {
                \nl \texttt{cross\_coded\_pkts = encode(q)}\\
                \nl \texttt{send(dc2\_id, cross\_coded\_pkts)}\\
            }
            \nl \Else {
                \nl \texttt{q.clear()}\\
            }
            \nl \texttt{\textbf{break}}\\
        }    
    }
    
    \BlankLine

    \nl \texttt{q.push(pkt)}\\
    \nl \If {\texttt{q.isFull()}} {
        \nl \texttt{cross\_coded\_pkts = encode(q)}\\
        \nl \texttt{send(dc2\_id, cross\_coded\_pkts)}\\
    }
 \caption{Coding algorithm at DC1.}
\label{alg:dc1}
\end{algorithm}
\paragraph{Coding Algorithm.}
DC1 follows Algorithm~\ref{alg:dc1}, which captures the task of encoding across multiple flows at once. DC1 maintains two sets of queues: one set for in-stream encoding (one set per flow),
and a set for cross-stream encoding (one set per $k$).
When a packet arrives, it is copied and pushed into one queue of each type. 

Lines 1-5 check whether the relevant in-stream queue has reached a threshold, and if so, create coded packets and send them to DC2. For cross-stream coding, DC1 first selects the set of queues destined for the same DC2, and then chooses the individual queue in round-robin order (lines 6-8). DC1 avoids placing multiple packets from the same flow in the same cross-stream queue; if there already exist packets from the same flow in all queues, then DC1 processes the oldest queue. If there is only a packet from the flow in question, then the old packet is evicted and discarded, since sending cross-stream packets with only packets from a single stream reduces its effectiveness (lines 9-19). Once the packet is pushed into a cross-stream queue, if a threshold is reached, then coded packets are generated and sent to DC2 (lines 20-23).



Timing constraints pose a challenge to this algorithm. If one flow is much faster than all other flows, DC1 cannot hold back recovery data from the faster flow to wait to make full recovery packets. Therefore, we create a timer for each in-stream and cross-stream queue (not shown in Algorithm~\ref{alg:dc1}). On expiry of a queue timer, DC1 encodes all packets in the queue and sends them to DC2. 

\subsection{Recovery Protocol}
\label{subsec:loss-recovery}
\sys{} uses a receiver-driven recovery protocol: the onus is on the receiver to quickly \emph{detect} packet loss and undertake \emph{recovery} with the help of DC2. 




\paragraph{Loss Detection.}
The key challenge in loss detection is how to make a fast, accurate prediction of whether a packet is lost (and thus needs to be recovered using the nearby DC). 
Note that our receiver based loss detection cannot use traditional sender-based timeout mechanisms (e.g., TCP RTO) because the receiver does not have a notion of \emph{when} a particular packet is sent by the sender. 

In \sys{}, the receiver detects a loss if
either a gap in sequence numbers is detected (the simple case) or a timer expires for the next expected packet. Setting a suitable timeout value -- low enough for fast recovery, but high enough not to cause spurious timeouts -- requires learning and predicting packet arrival times. While this opens up the possibilities to use machine learning algorithms, our current design uses a simple two-state Markov model that works well for our workloads.

\emph{Two-State Markov Model.} The model utilizes packet inter-arrival time probabilities (based on past packet arrival history) to switch between two timeout values: it uses a \emph{small} timeout value for packets arriving within a burst (i.e., sub-RTT scale), and a \emph{long} timeout value across packet bursts or application sessions. Initially, the receiver starts off with the long timeout value, which is a function of the RTT of the path, but as soon as it starts receiving packets with a short inter-arrival time, it switches to the small timeout value -- a value chosen based on previously observed inter-arrival times of packets within a burst. It remains in this state until the small timeout expires and switches immediately to the long timeout value after sending a NACK.  To avoid spurious recoveries at burst or session boundaries -- indicated by a NACK arriving before the corresponding coded packet -- DC2 first checks with the receiver before undertaking the recovery. 

\subsubsection{Loss Recovery Overview}

Upon receiving a NACK from the receiver, DC2 must determine which loss recovery scheme is best: in-stream or cross-stream cooperative recovery.  Using a cheapest first philosophy, 
in-stream recovery from coded packets already present at DC2 is preferred. In this process, DC2 forwards the relevant in-stream coding packets to the receiver, who recovers lost data by decoding them in combination with other data packets from the stream.

When immediate in-stream recovery is inadequate, the question of which loss recovery mechanism to pursue presents an interesting tradeoff.  If the time budget allows for it, it may be tempting to wait
to see if more in-stream coded packets or ACKs arrive at DC2, signaling that in-stream recovery is sufficient, as this is cheaper in terms of both cost (bits transmitted) and recovery. However, cooperative recovery has two advantages. Coding across streams enables us to recover from bursty losses or outages in addition to random losses, and also allows the recovery process to begin \emph{immediately}, since the data packets used for cross-stream coding (and needed for decoding) tend to have arrived at other receivers in close temporal proximity to the losses, by construction. Thus, we prioritize cross-stream decoding whenever in-stream recovery is insufficient at decision time.

\paragraph{Fast Cooperative Recovery.} As an optimization, we make packet recovery faster by  using a proactive loss detection mode. In this mode, if DC2 receives enough consecutive loss recovery requests (NACKs) from a receiver, it does not wait for additional NACKs, and instead proactively initiates cooperative recovery for subsequent packets in anticipation that they may also be lost. DC2 switches back to normal recovery upon receiving an ACK, once the receiver starts receiving packets on the direct path from the sender again.

\paragraph{Cooperative Recovery Algorithm.}

 \begin{figure}[!t]
\centering
 \includegraphics[scale=0.14]{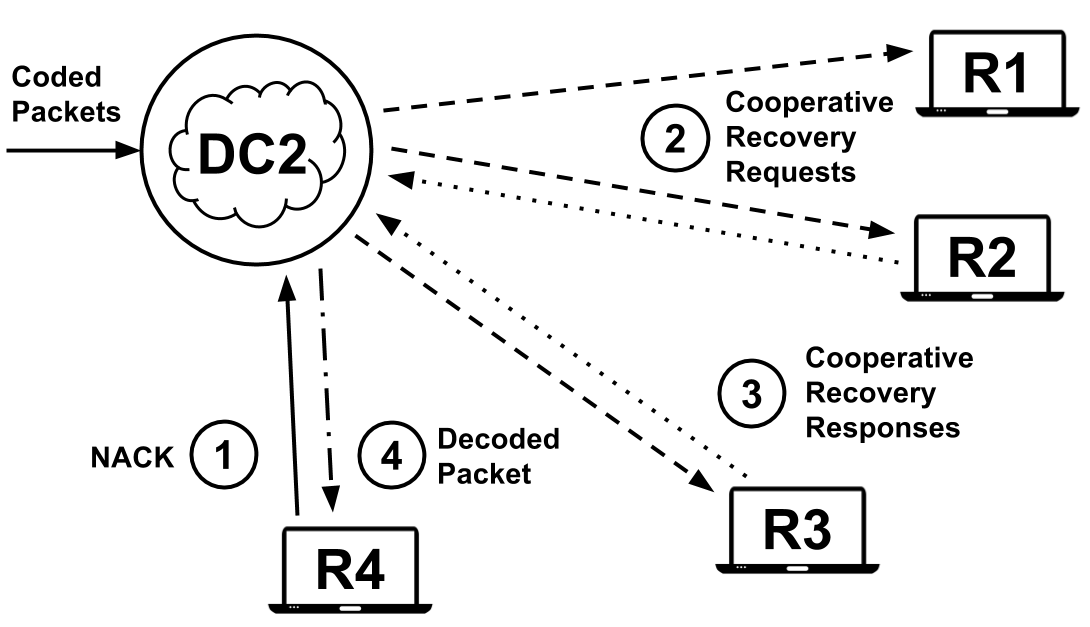}
 \caption{Cooperative Recovery.
 }
 \label{fig:coop_recovery}
 \end{figure}

Figure~\ref{fig:coop_recovery} shows a detailed look at the cooperative recovery protocol.
After receiving a NACK from receiver R4 (step 1), DC2 decides which type of recovery to use. After choosing cooperative recovery, it checks that there are sufficient cross-stream coded packets to conduct the recovery process. If so, DC2 parses each one to retrieve the addresses of the relevant receivers, and then sends cooperative requests to those receivers since they have the data packets needed to decode the missing packets (step 2). 
%
DC2 then processes any incoming cooperative recovery responses from the solicited receivers (step 3).
%
By tracking responses, DC2 can tabulate the number of cooperative responses for each recovery event. 
For each loss, once the number of responses is equal to $k - 1$ 
then recovery is possible. DC2 then decodes the lost packets and sends them to the receiver (4). Depending on the number of cross-stream coded packets, DC2 may only require a few of the receivers to respond in a timely fashion, thereby ignoring stragglers (such as $R1$ in Figure~\ref{fig:coop_recovery}) that can cause delay in recovery.
Since recovery is time sensitive, the protocol fails silently if not enough coded packets or cooperative recovery responses are received within a set deadline. We discuss these conditions under which recovery is not possible in Section~\ref{sec:wide_area_results}

Repeatedly applying this cooperative recovery process not only enables \sys{} to recover an \emph{indefinite} series of losses in most circumstances, but also makes the recovery process relatively fast compared to wide area retransmissions. Since each cross-stream packet is encoded from a batch of data packets (across different flows) with similar send times, cooperative recovery can be performed immediately, as opposed to waiting to collect enough packets for decoding from a single, linear stream. This, in combination with the ability to tune the coding rate to mitigate the effects of stragglers, relieves the temporal constraints imposed by time budgets. Additionally, the acknowledgement and cooperative recovery process in \sys{} exploits the current cloud pricing model, in which all incoming traffic is free and outgoing traffic is charged~\cite{amazonprice,azureprice,googleprice}. Since the size of a cooperative request is small, each data center can initiate send requests to multiple destinations inexpensively and can receive the bulkier data packets free of charge.   We evaluate how fast we can recover using real world data in Section~\ref{sec:wide_area_results}.

\section{Implementation}
\label{sec:implementation}
We now describe how to coordinate clients and data centers in practice, and elaborate on the prototype that we built to evaluate \sys{}.

\subsection{Bootstrapping \sys{}}

\paragraph{Data Center Selection.}
Before any communication can take place in \sys{}, senders and receivers must learn about their nearest \sys{}-enabled data center. On boot, a client only has knowledge of a master data center or web service, which it contacts to determine the location of this nearest data center. Optimally, the ``nearest'' data center will be the one with the lowest latency to the inquiring client, since our goal is to provide timely packet recovery. 

\paragraph{Clients Joining \sys{}.}
Once a client is ready to send data using the \sys{} service, it uses a control channel to contact its nearest data center (DC1) to set up a cloud overlay path. DC1 must then select an acceptable receiver-side data center (DC2), which it can do using the same data center selection methods. The chosen DC2 node must be close to the receiver in order for recovery to work within a time budget.
%
 DC1 and DC2 will provision the necessary resources and alert the sender that it may begin sending. From that point, the sender can begin transmitting data to DC1 for \sys{} 
to use as input for encoding over the cloud overlay path. 

\subsection{\sys{} Prototype}

The \sys{} prototype is implemented in C++ and operates in user space. 
Our implementation uses UDP for forwarding application traffic, coded packets, and cooperative recovery packets, and uses TCP for control channel traffic between the endpoints and the data centers. Applications can utilize \sys{} in two ways. First, iptables~\cite{iptables} and the NetFilter library~\cite{netfilter} can be used to install forwarding rules designed to ``catch'' outbound application traffic, redirect it to \sys{} to duplicate it on the data center path -- we use this for closed-source applications like Skype, where we are not at liberty to modify packets before they are transmitted by the application. Alternatively, \sys{} can act as a proxy listening on a local port for applications to directly send data. The data is received and encapsulated in the \sys{} header before being sent to the destination and DC1.

A \sys{} packet is essentially a UDP packet with application data packet as payload. \sys{} has a basic 32 byte header containing
standard flow fields.
Based on the type of packet, the header may include extra information. For example, an application data packet will include only the base header, whereas a coded packet will include information about how many flows are encoded, which flows are encoded, and which sequence numbers are involved. Similarly, cooperative recovery request and response packets contain information about the requested and lost packets.

Our prototype uses Reed-Solomon codes to encode and decode application data using the open-source zfec~\cite{zfec} library, and uses 
25ms for the small timer and RTT for the long timer.
Proactive loss detection mode is triggered with a string of three consecutive NACKs, similar to the TCP best-practice of triple duplicate ACKs. Finally, we tune the parameters related to coding (coding rate, timers, and queues) on a per-application basis, depending on the application's characteristics and requirements.

\section{Evaluation}
\label{sec:evaluation}
We perform a multi-tiered evaluation with the goals of answering: 
 (1) How effectively does \sys{} recover packets within a time budget for wide-area paths (\S~\ref{subsec:pl-eval})?
 (2) How does \sys{} perform in the contexts of challenging application, transport, and network requirements (\S~\ref{subsec:skype-perf}, \ref{subsec:rewan-tcp-eval}, \ref{subsec:otherdeployment})?
 (3) How effectively does the \sys{} prototype scale to support the demands of many application streams, and what is the monetary and encoding cost of scaling up (\S~\ref{subsec:prototype-eval})?

 %
 %
  \begin{figure*}[!t]
\centering
  \subfigure[]{\includegraphics[width=1.37in]{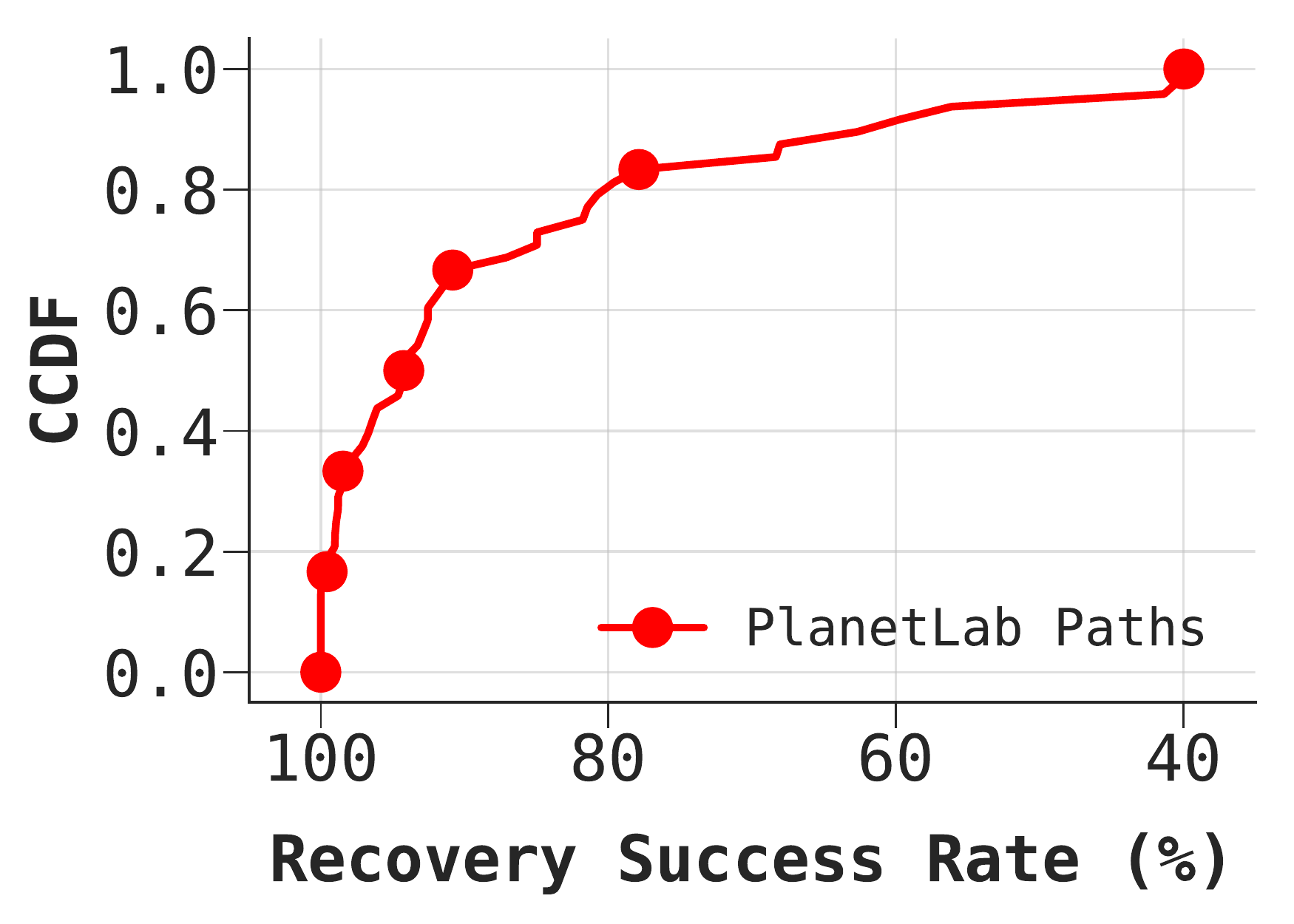}\label{fig:rewan_recovery_overview}}
  \subfigure[]{\includegraphics[width=1.37in]{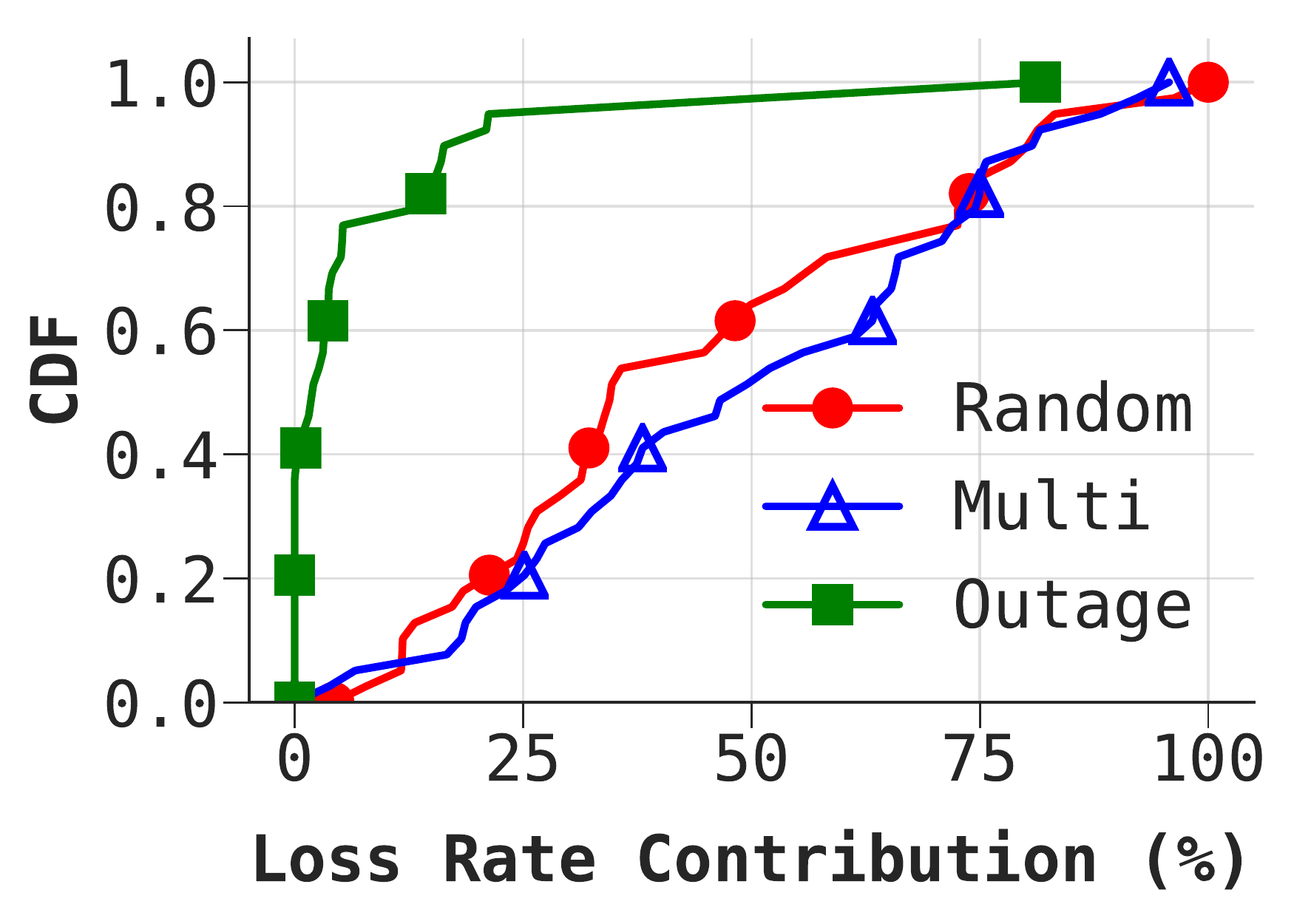}\label{fig:rewan_burst_overview}}
  \hspace{0.01em}
  \subfigure[]{\includegraphics[width=1.37in]{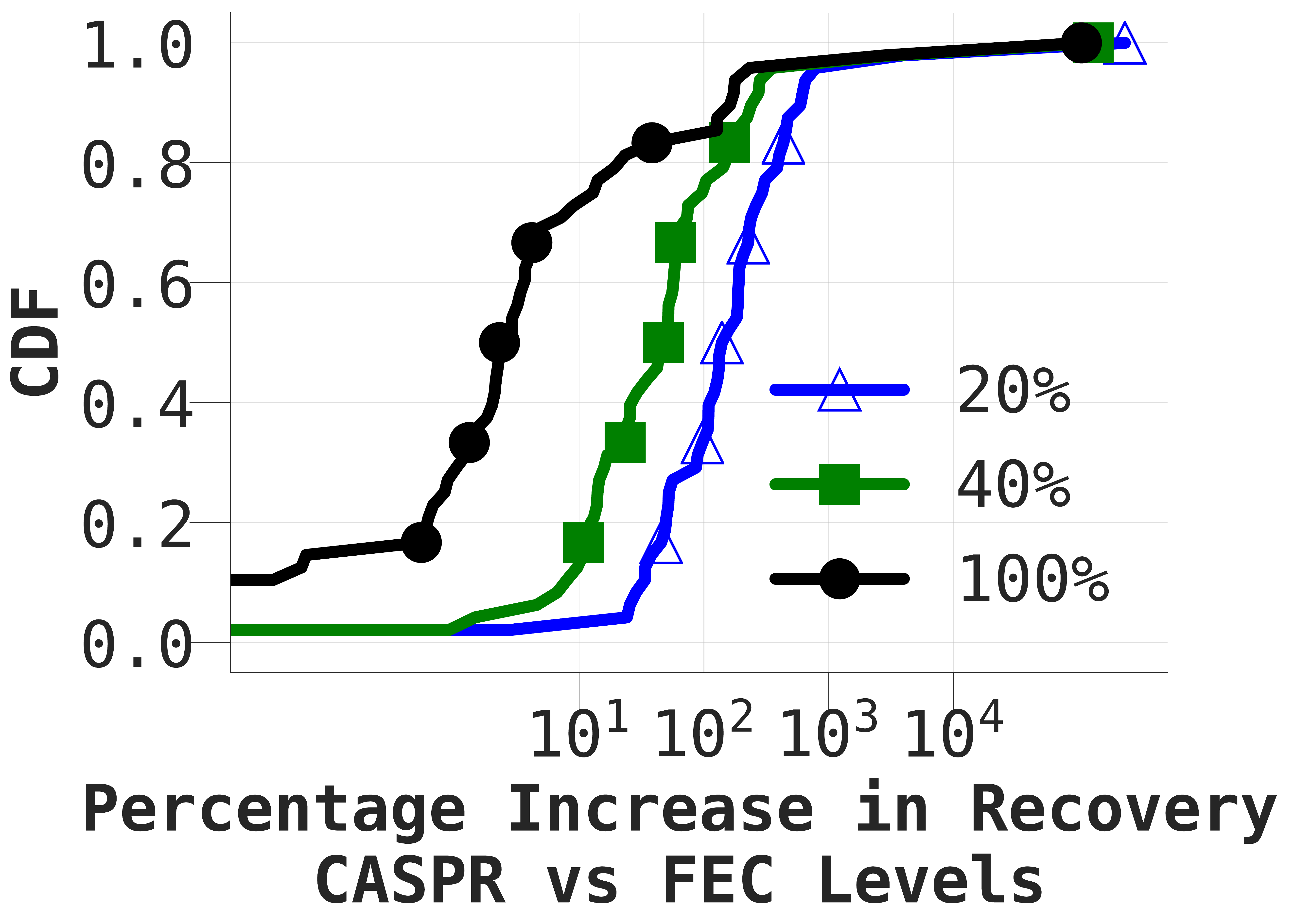}\label{fig:rewan_fec_diff}}
  \subfigure[]{\includegraphics[width=1.37in]{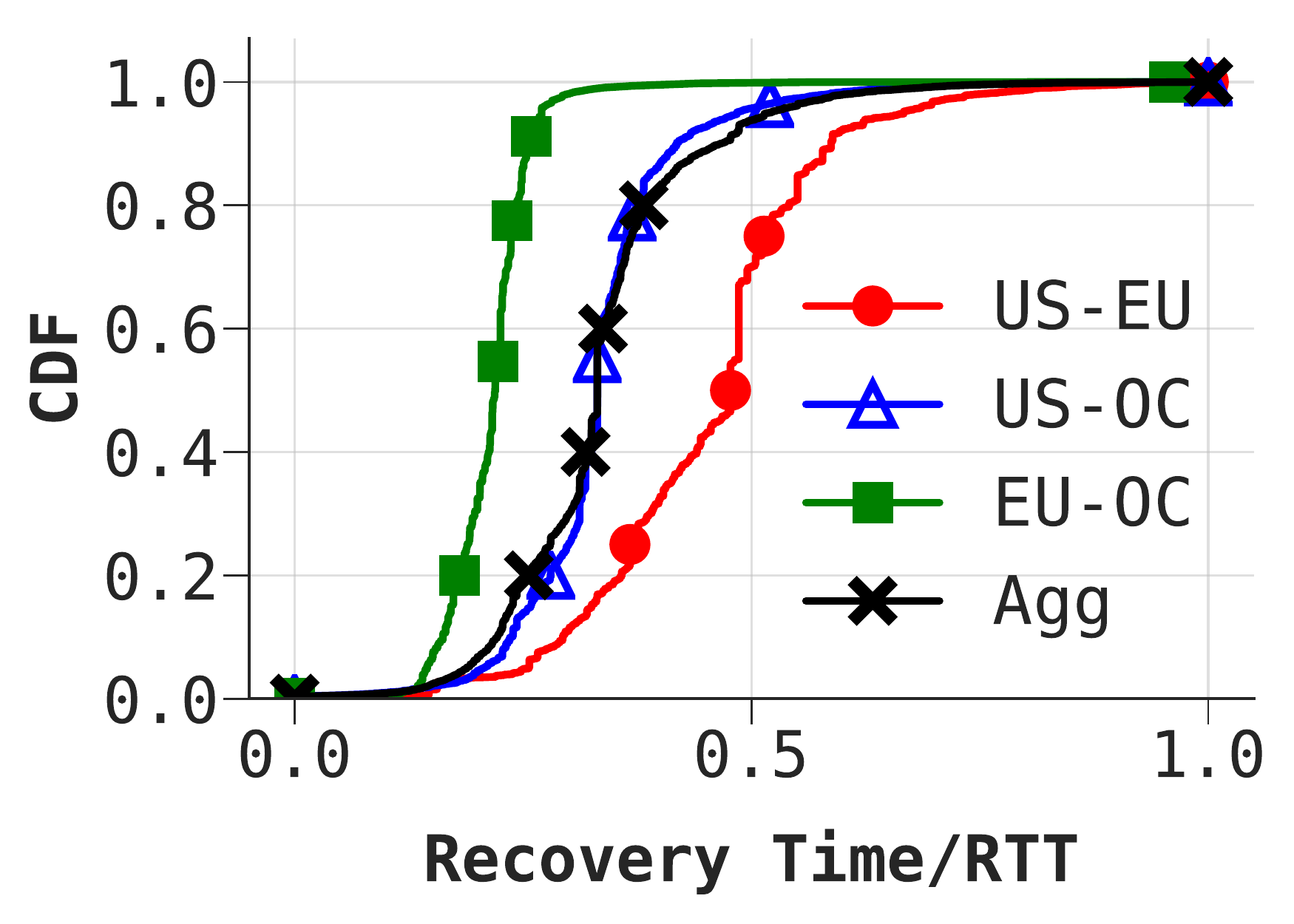}\label{fig:rtime_rtt}}
 \subfigure[]{\includegraphics[width=1.37in]{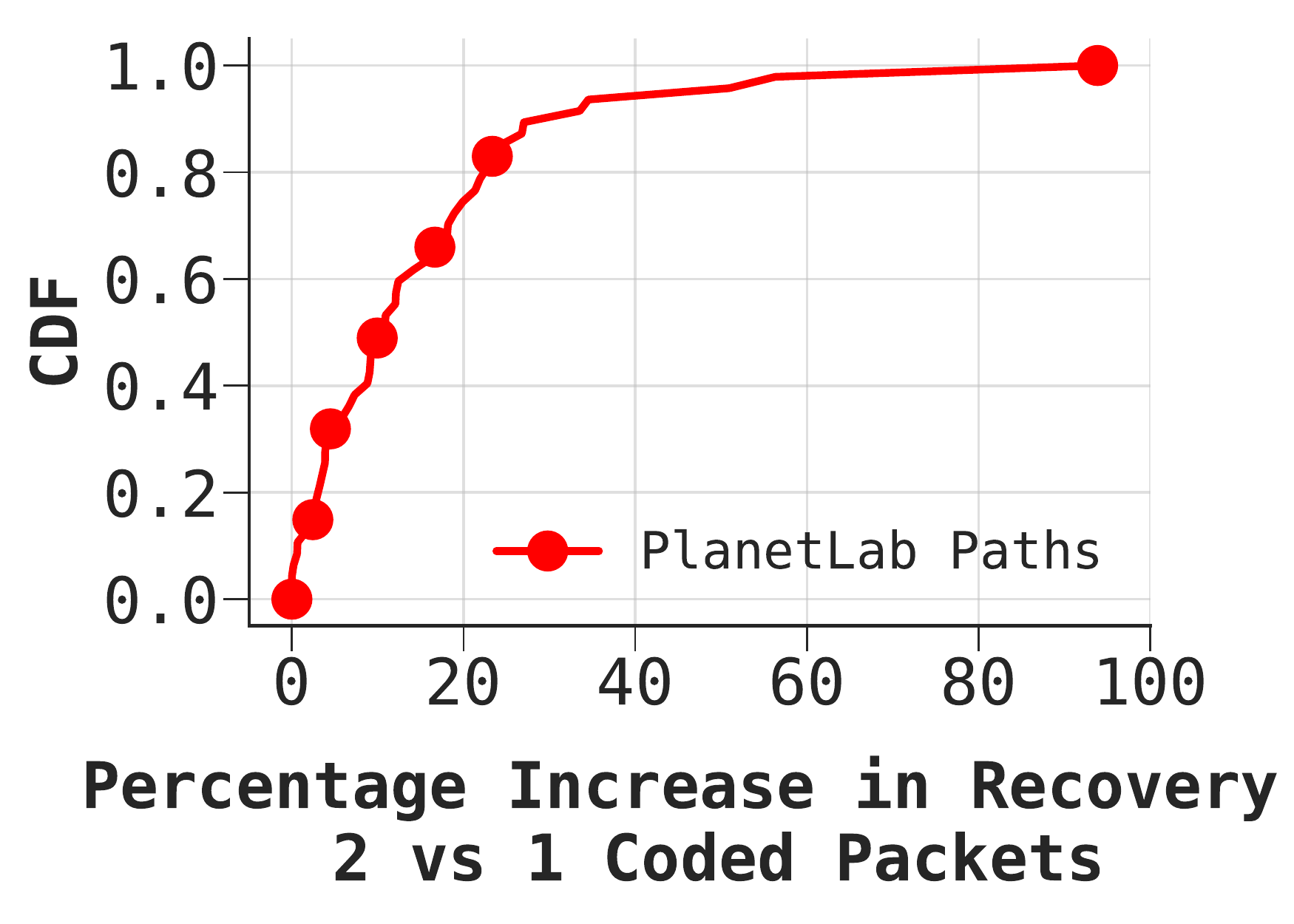}\label{fig:reco_diff}}

  \caption{\sys{}'s performance on PlanetLab paths.(a) CCDF of successfully recovered packets. (b) Loss episode contribution to loss rate on paths with greater than 80\% recovery. (c) Percentage increase in \sys{} recovery rate vs FEC with 20\%, 40\%,and 100\% packet overhead on direct path (note x-axis). (d) Packet recovery times as ratio of direct path RTT. (e) Percentage increase in recovery rates using 2 cross coded packets per batch versus 1. }
\label{fig:rewan-all-metrics}
 \end{figure*}
 
 Our evaluation encompasses: 

\noindent\textbf{Wide-area Deployment and Evaluation.} We ran \sys{} on a public cloud for over a month and measured its effectiveness in recovering losses for 45 inter-continental PlanetLab paths (\S\ref{subsec:pl-eval}). Our results show that \sys{} is able to recover 78\% of the packet losses. Further, in  over 80\% of cases, the recovery time is less than half an RTT of the direct path between the sender and receiver. 
     
\noindent\textbf{Cross-Layer, Controlled Environment Evaluations.}
     We evaluate \sys{} with respect to user-level QoE metrics for Skype video conferencing (\S\ref{subsec:skype-perf}), flow completion time for short TCP flows (\S\ref{subsec:rewan-tcp-eval}), network QoS in a cellular setting (\S\ref{subsec:mobile-eval}), and other duplication strategies (\S\ref{subsec:deployment}). We find that in controlled environments, \sys{} can provide benefits up and down the stack, albeit with some environment-specific tuning to keep it cost-effective. 
    
\noindent\textbf{Scalability and Cost Evaluation.} We evaluate the scalability of our prototype (\S\ref{subsec:prototype-eval}) -- our results show that a single thread can handle 150 concurrent video calls and we can get linear speed-up as we increase the number of \sys{} threads. We also show
that the cost of supporting a large number of user streams is low compared to a full overlay solution.

\subsection{Wide-Area Deployment and Evaluation}
\label{subsec:pl-eval}

\subsubsection{Setup} 
We have been running \sys{} as a service on five different DCs of Microsoft Azure~\cite{azure}, located in US, EU, Asia, and OC, for over a month. We use F1 type virtual machine, which is compute-optimized with 2.4 GHz single core and 2 GiB RAM. 
We evaluate 45 PlanetLab wide area paths spanning four different continents\footnote{Exact path details can be found at \url{http://tinyurl.com/pl-paths}}.



We run a simple constant bitrate application on the PlanetLab nodes. 
To observe long-term time-averaged behavior without overloading the paths, we use ON/OFF periods with Poisson OFF times and constant ON times. In each ON interval, we send packets for 5 minutes; we set the mean OFF time to be 55 minutes.
DC1 relays the start of each ON interval to senders using a separate control channel, thereby ensuring that 
senders are (loosely) synchronized. We use $r=2/6$ and $s=1/5$ as our coding parameters.
Given the high churn rate of PlanetLab nodes, the total samples collected from each path varies.  Typically, we recover 500-800 samples per path, which translates to 3-5 weeks of measurement collection.

\subsubsection{Results}
\label{sec:wide_area_results}
Our wide-area evaluation makes five key findings, summarized below, and visually in 
Figure~\ref{fig:rewan-all-metrics}.

\paragraph{Most losses happen on wide-area links and \sys{} is able to recover them.} 
\sys{} is able to recover 78\% of all packets that are lost on the PlanetLab paths. Loss rates on these paths are relatively high: up to 0.9\% loss, with 40\% of paths having a loss rate greater than 0.1\%.  Overall, we lose 0.02\% packets in our experiment and we consider any packet that takes longer than one RTT to recover as a lost packet. As we discuss later, most of the packets that \sys{} is unable to recover are lost on the access paths. If we ignore those losses, \sys{}'s packet recovery goes up significantly. 
Figure~\ref{fig:rewan_recovery_overview} elaborates on the above results -- it shows a CCDF of the fraction of successfully recovered packets (i.e., those lost packets that are recovered within one RTT) for all PlanetLab paths. Most paths experience high recovery (low unrecovered packet rate) -- overall, 82\% of paths successfully recover more than 80\% of lost packets.

\paragraph{\sys{}'s coding is able to handle a wide range of loss patterns.} We next zoom into the loss patterns to understand what types of losses are being recovered by \sys{}. Figure~\ref{fig:rewan_burst_overview} shows a CDF of loss episode patterns observed on PlanetLab paths that have greater than 80\% packet recovery (82\% of total paths). We look at the burst length of the loss episode and classify them as Random (single packet loss), Multi-Packet (2-14 packets), and Outage ($>$14 packets). We observe all three types of loss patterns on the chosen paths. While random and multi-packet bursts contribute more towards the loss rate, outages are not uncommon on these paths. Our data shows that 45\% of paths see outages that last from 1 to 3 seconds. Our recovery rates show that \sys{} is able to handle multiple types of burst lengths, quickly.

\paragraph{Most access losses can be recovered using existing techniques.} 
While access losses (between source-DC1 and DC2-receiver) are not the main focus of \sys{}, we look at their loss characteristics to see whether well-known techniques can be used to recover such losses.  Our results show that around 98\% of such losses occur on source-DC1 paths and that a significant fraction, 90\%, of loss bursts are single packet losses and can be recovered using simple retransmissions (ARQ) or other simple redundancy based techniques (e.g.,~\cite{gentleaggression}) at the edges (i.e., between the end-points and the DCs).  In future, we plan to augment \sys{} to incorporate this observation.

\paragraph{\sys{} vs. On-Path FEC schemes.}
To compare \sys{} with traditional, on-path FEC packet recovery schemes, we perform a what-if analysis on the probes sent on the direct PlanetLab paths.
Our goal is to compare \sys{} with sending different number of FEC packets on the direct path. 
We divide the probes into 5 packet bursts and consider the next burst as the FEC packets. 
We then compute recovery success rates for 20\% ($s=\frac{1}{5}$), 40\% ($s=\frac{2}{5}$), and 100\% ($s=\frac{5}{5}$) FEC overhead. We also assume that, for \sys{}, access losses can be recovered using existing ARQ-based techniques. 

Figure~\ref{fig:rewan_fec_diff} shows the percentage increase in recovery rates for all the paths using \sys{}, compared to different levels of FEC. We observe that, even at 100\% overhead (full duplication), 90\% of the paths had at least one loss episode
that could have been recovered using \sys{} but not with on-path, 100\% FEC overhead. 
Further, 10\% of the paths observe more than 160\% improvement in recovery rates with \sys{} compared
to full, on-path duplication. These are paths that experience long burst of losses or outages that cannot be recovered using FEC on the direct path.  For 20\% overhead scheme, 100\% increase in recovery rate is seen by 70\% of the paths. This result shows that there exist paths for which \sys{}'s cross-stream coding is more effective in recovering from outages and bursty losses compared to traditional, on-path FEC based schemes.

\paragraph{\sys{}'s loss recovery is usually fast.}  We next look at packet recovery time using \sys{}, which  Figure~\ref{fig:rtime_rtt} depicts for paths in different regions. We show our recovery times as a ratio of direct public Internet path RTT between the source and destination. 
We note that 95\% of packets are recovered within 0.5 $\times$ RTT. As expected, we observe faster recovery for paths with higher absolute latency on the direct public Internet path. For example, on low RTT paths between the US and EU (110-130 ms), we see higher recovery times as a proportion of RTT, but in terms of absolute latency, 90\% of packets are retrieved within 75 ms. We also observe that receiver-DC2 RTTs on these paths vary significantly. For example, the RTT between receivers in the EU and their nearest data center varies from 16-70 ms ($\mu$ = 28 ms).
However, as cloud providers continue to strive towards reducing their latency to end-users~\cite{mappinggoogle}, we expect \sys{} recovery times to continue to improve over time.

Finally, we observe two systematic reasons contributing to the tail in the recovery time (Figure~\ref{fig:rtime_rtt}): delay in detecting and recovering a loss (e.g., due to delayed NACKs) and delay in arrival of coded packets at DC2. Overall, the percentage of recovered packets that fall outside of a reasonable time budget value is low and only accounts for roughly 1\% of the recovered packets.

\paragraph{Recovery time is improved due to straggler protection.}
Last, we show the benefit of using extra cross-stream coded packets to provide protection against stragglers during cooperative recovery. Figure~\ref{fig:reco_diff} 
shows the performance gains using two cross-stream coded packets per batch, as opposed to one. With adequate protection of two packets per batch, 60\% of paths see greater than 10\% improvement in recovery rates (Figure~\ref{fig:reco_diff}). 
We also observe that the recovery times decrease by at least 50 ms for 70\% of the recovered packets (not shown) -- in some instances, the difference is some stragglers that take several seconds.
This further justifies our choice of default parameter values for PlanetLab paths.

\subsection{Skype Performance with \sys{}}
\label{subsec:skype-perf}

We run \sys{} under Skype's video conferencing service to measure its effect on an interactive application.\footnote{Skype is transitioning some of its services from peer-to-peer to a cloud architecture, but that shift is incomplete as of submission time~\cite{skypecloud}; therefore, Skype's traditional peer-to-peer video conferencing architecture is used in this evaluation.}  We focus on the performance of Skype in wide-area settings where outages occur (similar to ones described earlier in our wide-area evaluation). To do so, we leverage the cloud path to run the video conference in three experiments. First, we examine how the video quality degrades during an outage along a public Internet path used by Skype. We then duplicate \textit{all} Skype packets over a cloud path to show that such a path can indeed make up for lost packets during outages. Finally, we use \sys{} to selectively transmit coded packets over the cloud path and perform recovery at the receiver.



\subsubsection{Testbed and Measurement Procedure} We use a similar testbed to that used by Zhang et al.~\cite{zhang2012profiling}, in which clients communicate using Skype's video conferencing service. We connect clients running Skype for Linux 4.3 in a LAN, and emulate wide area path characteristics such as latency, packet loss rate, and jitter.

We use Skype's screen sharing mode to transmit a pre-recorded video that closely represents the normal motions of human interaction during a video conference. We then compare the quality of each received video against the reference video by converting all videos to raw (uncompressed) format, and compute objective QoE scores on a frame-by-frame basis using VQMT~\cite{vqmt}. Although objective video quality metrics are not as reliable as subjective metrics given by users (such as Mean Opinion Score), they are sufficient to approximate the quality of the video on a frame-by-frame basis. We show the scores of each frame in a CDF to approximate the quality of each video in aggregate.

\subsubsection{Results}

\paragraph{Duplication using the cloud overlay enables higher QoE.} Figure~\ref{fig:rewan_skype_eval} shows the video quality results as we vary the network conditions and paths used. When a 30 second outage occurs along the Internet path, Skype's built-in FEC mechanism is insufficient to maintain an acceptable level of QoE. The video quality degrades with pixelation and frozen video, and the number of frames with poor PSNR scores significantly increases. Due to the high availability of the cloud path, when we duplicate all Skype data across the cloud path during the 30-second Internet path outage, virtually all packets reach the destination, preserving the video quality (similar to an Internet path with a 0\% loss rate). This shows that Skype is amenable to a packet recovery service running in tandem with it to correct losses on its direct public Internet path.

\paragraph{\sys{} achieves similar QoE compared to cloud duplication.} When running Skype over \sys{}, we disable in-stream coding on the cloud path ($s = 0$), since Skype uses its own FEC techniques on the Internet path to recover lost packets~\cite{chitchat}. To use cross-stream coding, we inject three \textasciitilde200 Kbps background UDP flows whose packets are coded with Skype packets at DC1 at a rate of $r = 1/4$, with $k = 4$. Figure~\ref{fig:rewan_skype_eval} shows that \sys{} achieves a similar level of QoE compared to duplicating across the cloud path. 

\paragraph{\sys{} uses significantly less bandwidth than cloud duplication.} Because Skype uses its own FEC, we only need to utilize cross-stream coding and recovery, reducing the amount of inter-DC bandwidth used. We also observed the inter-arrival time of packets during Skype calls, and tuned \sys{} accordingly by setting the NACK timeout value to 25 ms. This reduces the number of false positive NACKs that trigger unnecessary cooperative recovery. By taking advantage of this application-specific knowledge, \sys{} achieves similar QoE scores as full cloud duplication but uses much less bandwidth: in our experiments, \sys{} sent just 13.4\% as many packets and 13.6\% as many bytes as did the full utilization of the cloud overlay.



 \begin{figure}[!t]
  \subfigure[]{\includegraphics[width=2.05in]{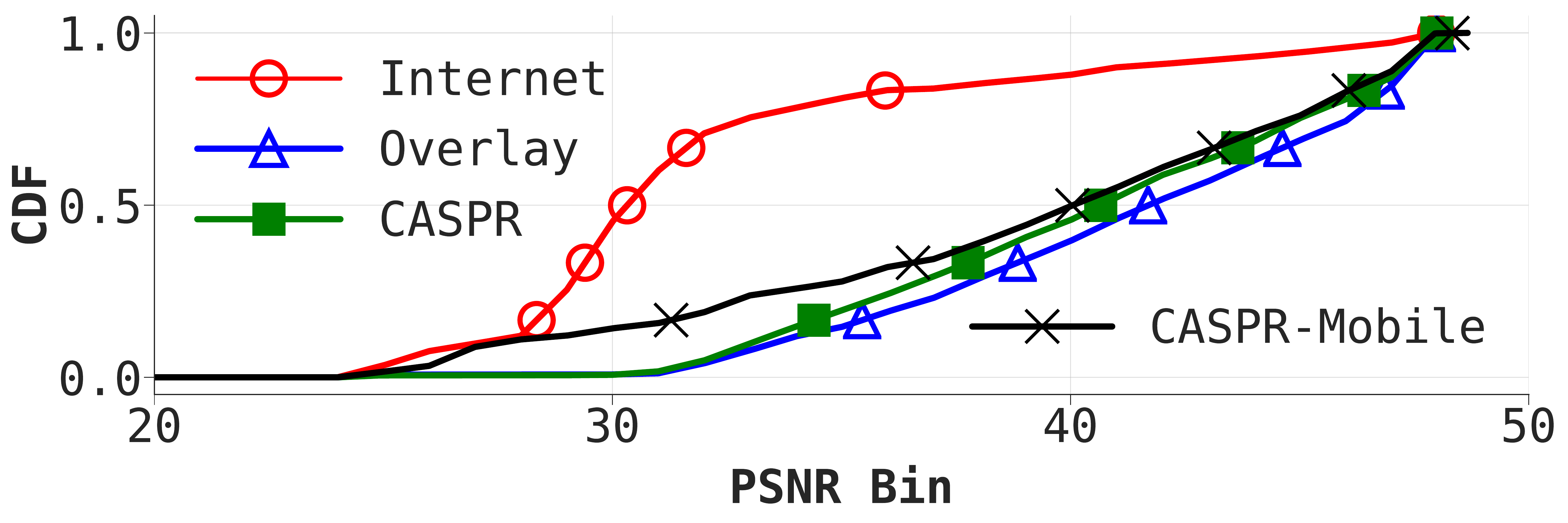}\label{fig:rewan_skype_eval}}
  \hspace{-0.60em}
  \subfigure[]{\includegraphics[width=1.05in]{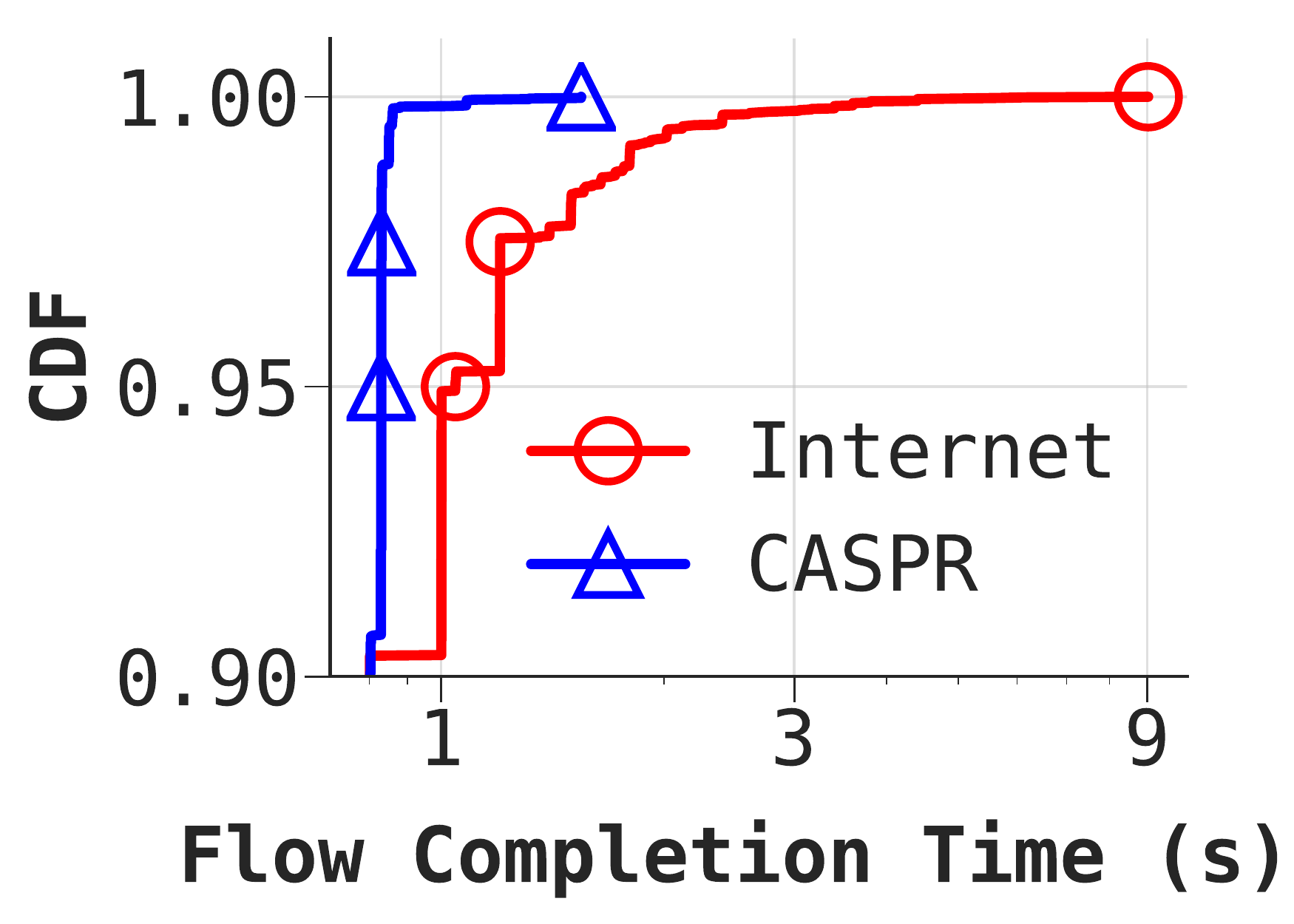}\label{fig:rewan_tcp}}
  \caption{a) PSNR scores of a set of video conferences,  b) Tail of TCP flow completion times (note y-axis scale).}
  \label{fig:rewan-s-and-t}
 \end{figure}
\subsection{TCP Performance with \sys{}}
\label{subsec:rewan-tcp-eval}
We now evaluate the performance of TCP if it is used over \sys{}. Our goal is to understand how the additional reliability provided by \sys{} interacts with TCP's own reliability and congestion control mechanisms, and whether it can provide any additional benefits. We focus on TCP short flows as they are latency sensitive and do not require high throughput.

\paragraph{Experimental Setup.} Our experimental setup is inspired by a similar experiment conducted by Google to evaluate different loss mitigation techniques for their web transfers~\cite{gentleaggression}. Using Emulab, we emulate the same topology and loss model as used in the Google study:  we consider a 200 ms RTT between end hosts and loss probabilities of 0.01 for losing the first packet in a burst and 0.5 for each subsequent loss. Given our focus on understanding TCP's interaction with \sys{}, we consider a single client-server scenario, in which a client sends a 12B request and receives a 50 KB response from the server. The RTT between server/client-DC paths is 30 ms with an RTT of 200 ms on the DC1-DC2 path. We make 10K requests each for TCP and TCP over \sys{}. 

\paragraph{\sys{} reduces tail latency for lossy short flows.}
Figure~\ref{fig:rewan_tcp} shows TCP's flow completion times with and without \sys{}. We observe that TCP suffers from long latency tail that goes up to 9 seconds, whereas \sys{} reduces the tail significantly. Our analysis shows that TCP is able to recover from most of the losses (using SACK), but there are some losses which are problematic for TCP, and hence cause the long tail. Such losses typically occur at the start of the connection, e.g., SYN-ACK(s), or at the very end. Such losses cause 
TCP to timeout, and successive losses mean that these timeout values could become huge, resulting in the long tail for TCP. 
\sys{} is able to reduce flow completion times by quickly recovering these losses. As soon as a packet is recovered by \sys{}, our TCP client sends an ACK to the server, effectively hiding the loss, and avoiding TCP timeouts. 





\paragraph{Two State Markov Model Reduces Overhead.} Note that TCP sender's control loop is quite different than the open loop CBR senders we have considered in our prior experiments. Our analysis also shows that our simple two state Markov model is able to adjust to TCP sender's control loop, specifically slow start, by using a smaller timeout in the middle of the window, and using a larger timeout across windows and subsequent transfers. Compared to maintaining a single timeout value, the two state approach results in 5x fewer NACKs sent to DC2. 





\subsection{Prototype Scalability and Cost}
\label{subsec:prototype-eval}


\noindent\textbf{Scalability.} We benchmark the performance of our prototype in terms of the number of encoded packets at DC1, since it is the most computationally expensive component of \sys{}. Our goal is to measure how efficiently \sys{} can process and encode packets as the system scales to a large number of concurrent streams. 
For each flow, \sys{} is configured to generate a single coded packet per every five data packets. We use Dell Poweredge R430 servers on Emulab, and each server is equipped with two 2.4 GHz 8-core processors with two threads each, for a total of 32 hardware threads.

\begin{figure}[!t]
 \centering
 \includegraphics[width=3.2in]{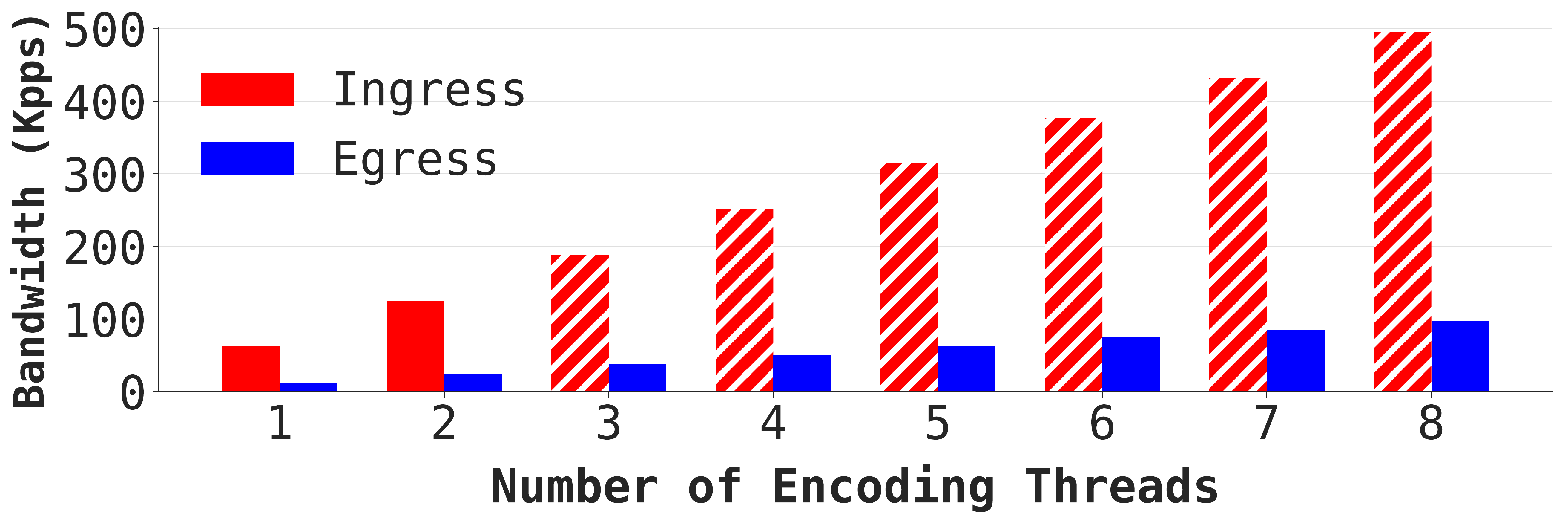}
 \caption{Throughput of the \sys{} prototype scales linearly with the number of encoding threads. Eight threads can handle up to 500Kpps.}
 \label{fig:proto_eval}
\end{figure}

We first determine the maximum throughput achievable at DC1 in packets per second. Measuring packets (instead of bits) is the appropriate measurement granularity because the encoder operates over entire packets. 
We find that a single encoding thread can handle around 65 Kpps. For context, assuming an average packet size of 512 bytes, 65 Kpps is enough for performing one-way processing for \textasciitilde150 simultaneous HD Skype video calls~\cite{skype_bw}. At this rate, the bottleneck is the generation of coded packets from data packets.

We then increase the number of DC1 encoding threads as we increase the number of senders, and load balance the streams to the different encoding threads. We rate limit each sender to 65 Kpps -- the empirical maximum rate that can be processed by a single (sender, encoder) pair. Figure~\ref{fig:proto_eval} shows that the processing power of \sys{} scales linearly with the number of encoding threads: up to \textasciitilde500 Kpps with eight encoding threads. 
This shows that \sys{} is amenable to parallelism and can be deployed in software to handle a large number of users. 


\noindent\textbf{Deployment Cost.} For estimating \sys{} deployment cost, we do a 
back-of-the-envelope calculation that compares \sys{} with a solution that fully uses the cloud. Based on the bandwidth requirement of Skype, a single user will send 0.675 GB of data per hour. Since a single \sys{} thread can handle 150 Skype calls, a data center node in an overlay will receive and forward $\sim$101 GB of data per hour, for 150 parallel application sessions. For a 2-node overlay node, based on today's cloud pricing~\cite{azureprice}, this would cost a minimum of \$17.60/hour for bandwidth and \$0.13/hour for single thread general purpose compute usage. However, for \sys{} using a coding rate of $r=1/16$, the \emph{maximum} cost of bandwidth for 150 calls will only be \$1.10/hour, which is 16x less than the cost of a full overlay. In this calculation, we are assuming that every coded packet will be used to recover a lost packet, which is an upper-bound on the overhead of \sys{} --  in practice, the outgoing bandwidth from DC2 will only be used in case of a packet loss.  


\noindent\textbf{Coding Overhead.} Due to limited availability of PlanetLab nodes, our wide area experiments considered a small number of concurrent user streams for encoding (typically $k=6$). 
We have also conducted experiments in Emulab to validate the feasibility of encoding over larger number of concurrent streams, which can further reduce the overhead of coded packets. 
Our controlled experiment, which used loss rates from Google's study (as in our TCP experiment), shows that for 20 concurrent streams and 2 cross stream coded packets (i.e., $r=2/20$), 
we can recover more than 92\% of the lost packets, for a coding overhead of only 10\%. 

\subsection{Other Deployment Scenarios}
\label{subsec:otherdeployment}
Our focus so far has been on scenarios with well connected endpoints. 
We now explore the potential use of \sys{}
in challenging scenarios, such as cellular access networks or 
when full duplication at source is infeasible. 

\subsubsection{Mobile Networks with \sys{}}
\label{subsec:mobile-eval}
The design of \sys{} makes assumptions that can be challenged in mobile networks, since mobile settings have different bandwidth, power, and latency characteristics.

\paragraph{Duplicating traffic can be feasible.} 
The bandwidth provided to cellular devices can vary greatly~\cite{sprout2013} -- our survey of major US carriers shows users can typically expect 2-5 Mbps uplink bandwidth. Therefore, we consider whether the most bandwidth intensive part of \sys{} -- the duplication of traffic to the cloud path at the sender -- works within the link rates of mobile networks.

We modified our Skype testbed (\S\ref{subsec:skype-perf}) to tether the sending host to a mobile device connected to an LTE network, and observed that the overall bandwidth required by \sys{} to duplicate a Skype video stream (1.5 Mbps) is well within the uplink bandwidth 
, which we measured to be $\sim$5.0 Mbps. However, the recommended bandwidth for HD video calls in Skype is 1.5 Mbps~\cite{skype_bw}, so duplicating that traffic to over 3.0 Mbps could reach the capacity of uplinks in some networks. We also tested how \sys{} affects other ongoing transfers on the device, and found that the transfer time for 5 MB files over WhatsApp is not affected by \sys{} running simultaneously.

For data-intensive uses, \sys{} may need to more selectively duplicate packets or only do so when the Internet path performance is below a certain requirement. 
Similarly, other deployment scenarios (\S\ref{subsec:deployment}) may become more relevant if the cellular provider employs a usage-based pricing model. 

%

%

\paragraph{Duplicating traffic has negligible impact on power consumption.} We tested the effect of duplicating a traffic stream on the battery life of the device. We ran 20 minute trials of Skype video calls, with and without cloud path duplication. We observed that in both cases the battery drain was $\sim$20 mAh, highlighting that the extra overhead of \sys{} has negligible impact on battery life. 


\paragraph{Recovery is feasible despite latency issues.} Mobile networks also suffer from greater end-to-end latency and jitter~\cite{sprout2013}. We conducted a short study to quantify this effect by pinging three major cloud providers (Amazon, Microsoft, and Google) 1,000 times using different mobile networks: Verizon's LTE network (east coast) and T-Mobile's LTE network (both east and west coasts). 
The median ping times to each provider was typically in the range of 50-60 ms, but the 50\%-90\% RTTs to each cloud provider was in the range of approximately 50-100 ms.

These latencies could be problematic for mobile receivers, as the effect of greater latency is multiplied in cooperative recovery. However, we observed that a Skype stream can be recovered during an outage,
using four mobile receivers, we see similar PSNR scores as fixed line (Figure~\ref{fig:rewan_skype_eval}). This is because Skype is able to adapt to greater end-to-end delay when recovery occurs over a longer duration.
 Due to increased jitter, correcting random packet losses may be difficult for interactive applications, but can likely be mitigated for other applications (such as web transfers) using in-stream coding.


\subsubsection{Duplication Strategies}
\label{subsec:deployment}

Our focus in this paper has been on scenarios where it is feasible for the sender 
to duplicate packets; we now explore strategies where this may not be feasible. 

\noindent\textbf{Selective Duplication.} When full duplication at source is infeasible -- due to limited access bandwidth or applications with high bitrates -- we can use \sys{} only for selected packets. To demonstrate the feasibility of such a strategy (and its potential benefits), we modify our TCP experiment (\S\ref{subsec:rewan-tcp-eval}) and only duplicate SYN-ACK packets. We observe that selective duplication reduces tail by 33\% (83\% with full duplication). Other examples of such duplication can include I-frames for video streaming, important user actions for gaming or AR applications, and the last packet of a window for short TCP transfers~\cite{gentleaggression}

\noindent\textbf{Duplication inside the network.} Another strategy, which requires support from the network, involves setting a bit in the packet header (e.g., TOS), which could indicates whether packet recovery is required or not. Based on this bit,
the ISP could duplicate the packet and send it to the cloud for packet recovery. Similarly, this duplication can also happen at a nearby data center, which can send the original packet through the public Internet and use the cloud path for coded packets only.
Finally, the network can also duplicate and send the copy to an alternate replica, if the other end-point application is replicated
across multiple sites/servers~\cite{rans-hotnets, tail_at_scale}.

\section{Related Work}
\label{sec:related-work}

\sys{} connects to and benefits from a large body of prior work. We comment on key pieces from the literature that our most relevant to our study.

\textbf{Overlay Networks.} Our work is inspired by \emph{overlay} networks that improve availability by using detour points, e.g., RON~\cite{ron}, one-hop source routing~\cite{detour}, Spines~\cite{spines}, etc. Our use of the cloud as an overlay creates unique opportunities and challenges. For example, we can do sub-RTT recovery, but to minimize cost, we have to send an additional small number of recovery packets. 
Individual aspects of \sys{}'s design also resonate with other overlay based solutions. 
For example, applying coding across users is similar to applying QoS across streams~\cite{overqos}.
Recently, there have been proposals that make the case for using cloud as an overlay to support interactive applications (VIA~\cite{via2016})
, TCP-based applications (CRONets~\cite{cronets2016}), as well as delay tolerant applications~\cite{ccr-slack}. 
Schemes like VIA improve performance by routing \emph{all} of a certain user's traffic through the overlay path. Our evaluation shows that \sys{} can achieve similar benefits while only sending a fraction of traffic over the expensive inter-DC paths. Finally ReWAN~\cite{rewanhotnets} provides a high level idea of using coded packets across the cloud paths for packet recovery. In this paper, we provide a complete design, including a tunable coding algorithm and recovery protocol, as well as deployment and evaluation under realistic network conditions as well as under controlled environments. 

\textbf{Inter-Data Center Networking.} Our work complements the large body of work on \emph{inter-data center} networking. This includes application of software defined networking (SDN) to such environments (e.g., SWAN~\cite{swan}, B4~\cite{swan}), as well as techniques that meet specific workload needs (e.g., application deadlines~\cite{eurosys15, tempus}). Similarly,  studies on inter-DC measurements~\cite{cloudcmp, eurosys15} have mainly focused on inter-DC \emph{bandwidth}. \sys{}'s use of inter-DC paths to send coded packets for recovery is complementary to these prior efforts. 

\textbf{Coding and Cooperative Recovery.}
Traditionally, network coding techniques have seen widest use in  the context of wireless networks~\cite{Katti06,networkcoding2006}. 
\sys{} applies cross-stream coding on wide area Internet paths and uses it to recover lost packets.
FEC based coding schemes have also been used in different contexts over the last several decades. The most relevant work to our scheme is Maelstrom~\cite{maelstrom}, which uses an FEC-based technique to reduce packet loss on lambda networks. Maelstrom's layered interleaving provides additional protection against bursty losses, but at the expense of higher decoding delay, which limits its use for highly interactive applications. Also, unlike Maelstrom,  the coded and data packets are sent on \emph{different} paths, with very different properties.





\textbf{Reliable and Low Latency Wide Area Communication.} Finally, we share the goals of proposals that call for \emph{low latency and high reliability} for wide area communication~\cite{speedoflight, arrow, caip, npp}. For example, Arrow~\cite{arrow} is proposed as a reliable, wide area service;  it uses reliable paths as tunnels to improve end-to-end reliability. While inter-DC paths are likely to have similar properties as Arrow's reliable paths, our approach of only using these paths for recovery is different and complementary to Arrow's goals. 



\section{Discussion and Future Work}
\label{sec:discussion}

Our research opens up interesting future work, such as: 
  \paragraph {Privacy and security considerations.} Duplicating traffic along cloud paths to be processed in data centers opens up the possibility of surveillance. At least one United States governmental program has collected traffic between data centers abroad~\cite{arnbak2014loopholes}. Future work could also study the privacy implications of network coding across multiple streams, and incorporate better privacy mechanisms in \sys{}.
  
    \paragraph {Interface between \sys{} and the applications.} 
    \sys{}'s design involves various trade-offs involving performance and cost. We believe that ultimately the application (or the user) is in the best position to make suitable trade-off decisions, keeping in view its performance requirements as well as cost budget. Thus, exposing these trade-offs to the application through a suitable interface (e.g., through ChoiceNet~\cite{choicenet} or XIA~\cite{xiansdi}) is another promising research direction.




\section{Conclusion}
\label{sec:conclusion}

\sys{} seeks to connect two complementary interests: the \emph{pull} of existing (and burgeoning) applications and their demand for better user experience, and the \emph{push} of DC technology that makes cloud services more accessible to the edge than ever before. 
The key idea behind \sys{} is to use the cloud paths \emph{only} for recovery by sending coded packets across the reliable but expensive inter-DC paths. As more applications shift their services to the cloud and as highly interactive VR/AR applications continue to emerge, classical algorithmic techniques such as coding can address fresh challenges. We view \sys{} as a promising step toward providing application and network architects with new insights into how to judiciously leverage the cloud.

\newpage






\section*{Acknowledgments}
We thank the anonymous reviewers for their feedback on this work, and Shawn Doughty for helping with PlanetLab access. This work was partially supported by NSF CNS under award numbers 1815046 and 1815016.

\setlength{\bibsep}{2pt plus 1pt}  
\small
\bibliography{ref}

\begin{thebibliography}{57}
\providecommand{\natexlab}[1]{#1}
\providecommand{\url}[1]{\texttt{#1}}
\expandafter\ifx\csname urlstyle\endcsname\relax
  \providecommand{\doi}[1]{doi: #1}\else
  \providecommand{\doi}{doi: \begingroup \urlstyle{rm}\Url}\fi

\bibitem[ama()]{amazonprice}
{Amazon AWS Pricing}.
\newblock \url{https://aws.amazon.com/pricing/services/}.

\bibitem[aws()]{aws-direct-connect}
{AWS Direct Connect}.
\newblock \url{https://aws.amazon.com/directconnect/}.

\bibitem[azu({\natexlab{a}})]{azure}
{Microsoft Azure}.
\newblock \url{http://azure.microsoft.com/}, {\natexlab{a}}.

\bibitem[azu({\natexlab{b}})]{azure-expressroute}
{Microsoft Azure ExpressRoute}.
\newblock \url{https://azure.microsoft.com/en-us/services/expressroute/},
  {\natexlab{b}}.

\bibitem[azu({\natexlab{c}})]{azureprice}
{Microsoft Azure Pricing}.
\newblock \url{https://azure.microsoft.com/en-us/pricing/details/bandwidth/},
  {\natexlab{c}}.

\bibitem[goo()]{googleprice}
{Google Cloud Pricing}.
\newblock \url{https://cloud.google.com/interconnect/docs#pricing}.

\bibitem[ipt()]{iptables}
{netfilter iptables}.
\newblock \url{https://www.netfilter.org/}.

\bibitem[sky({\natexlab{a}})]{skype_bw}
{How much bandwidth does Skype need?}
\newblock \url{
  https://support.skype.com/en/faq/FA1417/how-much-bandwidth-does-skype-need},
  {\natexlab{a}}.

\bibitem[sky({\natexlab{b}})]{skypecloud}
{Skype – the journey we’ve been on}.
\newblock
  \url{https://blogs.skype.com/news/2016/07/20/skype-the-journey-weve-been-on},
  {\natexlab{b}}.

\bibitem[Amir and Danilov(2003)]{spines}
Y.~Amir and C.~Danilov.
\newblock Reliable communication in overlay networks.
\newblock In \emph{Proc. IEEE DSN}, 2003.

\bibitem[Andersen et~al.(2001)Andersen, Balakrishnan, Kaashoek, and
  Morris]{ron}
D.~G. Andersen, H.~Balakrishnan, M.~F. Kaashoek, and R.~Morris.
\newblock {Resilient Overlay Networks}.
\newblock In \emph{Proc. ACM {SOSP}}, 2001.

\bibitem[Arnbak and Goldberg(2014)]{arnbak2014loopholes}
A.~Arnbak and S.~Goldberg.
\newblock Loopholes for circumventing the constitution: Unrestricted bulk
  surveillance on americans by collecting network traffic abroad.
\newblock \emph{Mich. Telecomm. \& Tech. L. Rev.}, 21, 2014.

\bibitem[Aslam et~al.(2004)Aslam, Raza, Dogar, Ahmad, and Uzmi]{npp}
F.~Aslam, S.~Raza, F.~R. Dogar, I.~U. Ahmad, and Z.~A. Uzmi.
\newblock Npp: A facility based computation framework for restoration routing
  using aggregate link usage information.
\newblock In \emph{International workshop on quality of service in multiservice
  IP networks}, 2004.

\bibitem[Bakre and Badrinath(1997)]{Bakre97}
A.~V. Bakre and B.~Badrinath.
\newblock {Implementation and Performance Evaluation of Indirect TCP}.
\newblock \emph{IEEE Transactions on Computers}, 46\penalty0 (3):\penalty0
  260--278, 1997.

\bibitem[Balakrishnan et~al.(2008)Balakrishnan, Marian, Birman, Weatherspoon,
  and Vollset]{maelstrom}
M.~Balakrishnan, T.~Marian, K.~Birman, H.~Weatherspoon, and E.~Vollset.
\newblock {Maelstrom: Transparent Error Correction for Lambda Networks}.
\newblock In \emph{Proc. USENIX NSDI}, 2008.

\bibitem[Cai et~al.(2016)Cai, Le, Sun, Xie, Jamjoom, and Campbell]{cronets2016}
C.~X. Cai, F.~Le, X.~Sun, G.~G. Xie, H.~Jamjoom, and R.~H. Campbell.
\newblock Cronets: Cloud-routed overlay networks.
\newblock In \emph{Proc. ICDCS}, 2016.

\bibitem[Calder et~al.(2013)Calder, Fan, Hu, Katz-Bassett, Heidemann, and
  Govindan]{mappinggoogle}
M.~Calder, X.~Fan, Z.~Hu, E.~Katz-Bassett, J.~Heidemann, and R.~Govindan.
\newblock Mapping the expansion of google's serving infrastructure.
\newblock In \emph{Proc. ACM IMC}, 2013.

\bibitem[Chiu et~al.(2015)Chiu, Schlinker, Radhakrishnan, Katz-Bassett, and
  Govindan]{goog-onehop}
Y.-C. Chiu, B.~Schlinker, A.~B. Radhakrishnan, E.~Katz-Bassett, and
  R.~Govindan.
\newblock Are we one hop away from a better internet?
\newblock In \emph{Proc. IMC}, 2015.

\bibitem[Dean and Barroso(2013)]{tail_at_scale}
J.~Dean and L.~A. Barroso.
\newblock The tail at scale.
\newblock \emph{Commun. ACM}, 56\penalty0 (2):\penalty0 74--80, Feb. 2013.

\bibitem[Dobrian et~al.(2011)Dobrian, Sekar, Awan, Stoica, Joseph, Ganjam,
  Zhan, and Zhang]{qoe2}
F.~Dobrian, V.~Sekar, A.~Awan, I.~Stoica, D.~Joseph, A.~Ganjam, J.~Zhan, and
  H.~Zhang.
\newblock Understanding the impact of video quality on user engagement.
\newblock \emph{Proc. ACM SIGCOMM}, 2011.

\bibitem[Dogar(2018)]{ccr-slack}
F.~R. Dogar.
\newblock Towards slack-aware networking.
\newblock \emph{ACM SIGCOMM Computer Communication Review}, 48\penalty0
  (2):\penalty0 24--30, 2018.

\bibitem[Dogar and Steenkiste(2008)]{DogarM208}
F.~R. Dogar and P.~Steenkiste.
\newblock M2: {U}sing {V}isible {M}iddleboxes to {S}erve {P}ro-active
  {M}obile-{H}osts.
\newblock In \emph{Proc. ACM SIGCOMM MobiArch}, 2008.

\bibitem[Dogar and Steenkiste(2012)]{tapaconext}
F.~R. Dogar and P.~Steenkiste.
\newblock {Architecting for Edge Diversity: Supporting Rich Services Over an
  Unbundled Transport}.
\newblock In \emph{Proc. ACM CoNext}, 2012.

\bibitem[Dogar et~al.(2005)Dogar, Uzmi, and Baqai]{caip}
F.~R. Dogar, Z.~A. Uzmi, and S.~M. Baqai.
\newblock Caip: a restoration routing architecture for diffserv aware mpls
  traffic engineering.
\newblock In \emph{Proc. IEEE DRCN 2005}, 2005.

\bibitem[Dogar et~al.(2010)Dogar, Steenkiste, and Papagiannaki]{catnap}
F.~R. Dogar, P.~Steenkiste, and K.~Papagiannaki.
\newblock {Catnap: Exploiting high bandwidth wireless interfaces to save energy
  for mobile devices}.
\newblock In \emph{Proc. ACM Mobi{S}ys}, 2010.

\bibitem[Flach et~al.(2013)Flach, Dukkipati, Terzis, Raghavan, Cardwell, Cheng,
  Jain, Hao, Katz-Bassett, and Govindan]{gentleaggression}
T.~Flach, N.~Dukkipati, A.~Terzis, B.~Raghavan, N.~Cardwell, Y.~Cheng, A.~Jain,
  S.~Hao, E.~Katz-Bassett, and R.~Govindan.
\newblock Reducing web latency: the virtue of gentle aggression.
\newblock In \emph{Proc. ACM SIGCOMM}, 2013.

\bibitem[Fragouli et~al.(2006)Fragouli, Le~Boudec, and
  Widmer]{networkcoding2006}
C.~Fragouli, J.-Y. Le~Boudec, and J.~Widmer.
\newblock Network coding: An instant primer.
\newblock \emph{SIGCOMM Comput. Commun. Rev.}, 36\penalty0 (1):\penalty0
  63--68, Jan. 2006.

\bibitem[Govindan et~al.(2016)Govindan, Minei, Kallahalla, Koley, and
  Vahdat]{goog-avail}
R.~Govindan, I.~Minei, M.~Kallahalla, B.~Koley, and A.~Vahdat.
\newblock Evolve or die: High-availability design principles drawn from googles
  network infrastructure.
\newblock In \emph{Proc. SIGCOMM}, 2016.

\bibitem[Gummadi et~al.(2004)Gummadi, Madhyastha, Gribble, Levy, Wetherall,
  et~al.]{detour}
P.~K. Gummadi, H.~V. Madhyastha, S.~D. Gribble, H.~M. Levy, D.~Wetherall,
  et~al.
\newblock {Improving the Reliability of Internet Paths with One-hop Source
  Routing}.
\newblock In \emph{Proc. USENIX OSDI}, 2004.

\bibitem[Han et~al.(2012)Han, Anand, Dogar, Li, Lim, Machado, Mukundan, Wu,
  Akella, Andersen, et~al.]{xiansdi}
D.~Han, A.~Anand, F.~R. Dogar, B.~Li, H.~Lim, M.~Machado, A.~Mukundan, W.~Wu,
  A.~Akella, D.~G. Andersen, et~al.
\newblock {XIA: Efficient Support for Evolvable Internetworking}.
\newblock In \emph{Proc. USENIX NSDI}, 2012.

\bibitem[Hanhart and Hahling(2013)]{vqmt}
P.~Hanhart and R.~Hahling.
\newblock Video quality measurement tool ({VQMT}), 2013.

\bibitem[Haq and Dogar(2015)]{rewanhotnets}
O.~Haq and F.~R. Dogar.
\newblock {Leveraging the Power of the Cloud for Reliable Wide Area
  Communication}.
\newblock In \emph{Proc. ACM Hotnets}, 2015.

\bibitem[Haq et~al.(2017)Haq, Raja, and Dogar]{cloudstudy-www}
O.~Haq, M.~Raja, and F.~R. Dogar.
\newblock Measuring and improving the reliability of wide-area cloud paths.
\newblock In \emph{Proc. WWW}, 2017.

\bibitem[Hong et~al.(2013)Hong, Kandula, Mahajan, Zhang, Gill, Nanduri, and
  Wattenhofer]{swan}
C.-Y. Hong, S.~Kandula, R.~Mahajan, M.~Zhang, V.~Gill, M.~Nanduri, and
  R.~Wattenhofer.
\newblock {Achieving high utilization with software-driven WAN}.
\newblock In \emph{Proc. SIGCOMM}, 2013.

\bibitem[Iftikhar et~al.(2016)Iftikhar, Dogar, and Qazi]{rans-hotnets}
A.~M. Iftikhar, F.~Dogar, and I.~A. Qazi.
\newblock Towards a redundancy-aware network stack for data centers.
\newblock In \emph{Proc. HotNets}, 2016.

\bibitem[Jalaparti et~al.(2016)Jalaparti, Bliznets, Kandula, Lucier, and
  Menache]{interdcpricing2016}
V.~Jalaparti, I.~Bliznets, S.~Kandula, B.~Lucier, and I.~Menache.
\newblock Dynamic pricing and traffic engineering for timely inter-datacenter
  transfers.
\newblock In \emph{Proc. SIGCOMM}, 2016.

\bibitem[Jiang et~al.(2016)Jiang, Das, Ananthanarayanan, Chou, Padmanabhan,
  Sekar, Dominique, Goliszewski, Kukoleca, Vafin, and Zhang]{via2016}
J.~Jiang, R.~Das, G.~Ananthanarayanan, P.~A. Chou, V.~Padmanabhan, V.~Sekar,
  E.~Dominique, M.~Goliszewski, D.~Kukoleca, R.~Vafin, and H.~Zhang.
\newblock Via: Improving internet telephony call quality using predictive relay
  selection.
\newblock In \emph{ACM SIGCOMM}, 2016.

\bibitem[Kandula et~al.(2014)Kandula, Menache, Schwartz, and Babbula]{tempus}
S.~Kandula, I.~Menache, R.~Schwartz, and S.~R. Babbula.
\newblock Calendaring for wide area networks.
\newblock In \emph{Proc. ACM SIGCOMM}, 2014.

\bibitem[Katti et~al.(2006)Katti, Rahul, Hu, Katabi, Medard, and
  Crowcroft]{Katti06}
S.~Katti, H.~Rahul, W.~Hu, D.~Katabi, M.~Medard, and J.~Crowcroft.
\newblock {XORs in the Air: Practical Wireless Network Coding}.
\newblock In \emph{Proc. SIGCOMM}, 2006.

\bibitem[Laoutaris et~al.(2011)Laoutaris, Sirivianos, Yang, and
  Rodriguez]{netstitcher2011}
N.~Laoutaris, M.~Sirivianos, X.~Yang, and P.~Rodriguez.
\newblock Inter-datacenter bulk transfers with netstitcher.
\newblock In \emph{Proc. SIGCOMM}, 2011.

\bibitem[Li et~al.(2010)Li, Yang, Kandula, and Zhang]{cloudcmp}
A.~Li, X.~Yang, S.~Kandula, and M.~Zhang.
\newblock Cloudcmp: comparing public cloud providers.
\newblock In \emph{Proc. ACM IMC}, 2010.

\bibitem[Luckie et~al.()Luckie, Dhamdhere, Clark, Huffaker, and
  claffy]{congestion-imc14}
M.~Luckie, A.~Dhamdhere, D.~Clark, B.~Huffaker, and k.~claffy.
\newblock Challenges in inferring internet interdomain congestion.
\newblock In \emph{Proceedings of the IMC 2014}.

\bibitem[Peter et~al.(2014)Peter, Javed, Zhang, Woos, Anderson, and
  Krishnamurthy]{arrow}
S.~Peter, U.~Javed, Q.~Zhang, D.~Woos, T.~Anderson, and A.~Krishnamurthy.
\newblock One tunnel is (often) enough.
\newblock In \emph{Proc. ACM SIGCOMM}, 2014.

\bibitem[Sekar et~al.(2012)Sekar, Egi, Ratnasamy, Reiter, and
  Shi]{combsekar2012}
V.~Sekar, N.~Egi, S.~Ratnasamy, M.~K. Reiter, and G.~Shi.
\newblock Design and implementation of a consolidated middlebox architecture.
\newblock In \emph{Proc. NSDI}, 2012.

\bibitem[Shafiq et~al.(2014)Shafiq, Erman, Ji, Liu, Pang, and Wang]{qoe1}
M.~Z. Shafiq, J.~Erman, L.~Ji, A.~X. Liu, J.~Pang, and J.~Wang.
\newblock Understanding the impact of network dynamics on mobile video user
  engagement.
\newblock In \emph{Proc. ACM Sigmetrics}, 2014.

\bibitem[Sherry et~al.(2012)Sherry, Hasan, Scott, Krishnamurthy, Ratnasamy, and
  Sekar]{middleboxcloud}
J.~Sherry, S.~Hasan, C.~Scott, A.~Krishnamurthy, S.~Ratnasamy, and V.~Sekar.
\newblock Making middleboxes someone else's problem: network processing as a
  cloud service.
\newblock In \emph{Proc. SIGCOMM}, 2012.

\bibitem[Sherry et~al.(2015)Sherry, Gao, Basu, Panda, Krishnamurthy, Maciocco,
  Manesh, Martins, Ratnasamy, Rizzo, and Shenker]{middleboxsherry}
J.~Sherry, P.~X. Gao, S.~Basu, A.~Panda, A.~Krishnamurthy, C.~Maciocco,
  M.~Manesh, J.~a. Martins, S.~Ratnasamy, L.~Rizzo, and S.~Shenker.
\newblock Rollback-recovery for middleboxes.
\newblock In \emph{Proc. SIGCOMM}, 2015.

\bibitem[Shokrollahi(2006)]{raptor}
A.~Shokrollahi.
\newblock Raptor codes.
\newblock \emph{IEEE Trans. Inf. Theor.}, 52\penalty0 (6), June 2006.

\bibitem[Singla et~al.(2014)Singla, Chandrasekaran, Godfrey, and
  Maggs]{speedoflight}
A.~Singla, B.~Chandrasekaran, P.~Godfrey, and B.~Maggs.
\newblock {The Internet at the Speed of Light}.
\newblock In \emph{Proc. ACM HotNets}, 2014.

\bibitem[Subramanian et~al.(2004)Subramanian, Stoica, Balakrishnan, and
  Katz]{overqos}
L.~Subramanian, I.~Stoica, H.~Balakrishnan, and R.~H. Katz.
\newblock {OverQoS: An Overlay Based Architecture for Enhancing Internet QoS.}
\newblock In \emph{Proc. USENIX NSDI}, 2004.

\bibitem[Wang(2010)]{chitchat}
J.~Wang.
\newblock \emph{ChitChat: Making video chat robust to packet loss}.
\newblock PhD thesis, Massachusetts Institute of Technology, 2010.

\bibitem[Welte and Ayuso()]{netfilter}
H.~Welte and P.~N. Ayuso.
\newblock Net{F}ilter.
\newblock \url{http://www.netfilter.org}.

\bibitem[Wilcox-O’Hearn(2008)]{zfec}
Z.~Wilcox-O’Hearn.
\newblock Zfec 1.4.
\newblock \emph{Open source code distribution: \url{http://pypi. python.
  org/pypi/zfec}}, 2008.

\bibitem[Winstein et~al.(2013)Winstein, Sivaraman, and
  Balakrishnan]{sprout2013}
K.~Winstein, A.~Sivaraman, and H.~Balakrishnan.
\newblock Stochastic forecasts achieve high throughput and low delay over
  cellular networks.
\newblock In \emph{{USENIX} {NSDI}}, 2013.

\bibitem[Wolf et~al.(2014)Wolf, Griffioen, Calvert, Dutta, Rouskas, Baldin, and
  Nagurney]{choicenet}
T.~Wolf, J.~Griffioen, K.~L. Calvert, R.~Dutta, G.~N. Rouskas, I.~Baldin, and
  A.~Nagurney.
\newblock Choicenet: {T}oward an {E}conomy {P}lane for the {I}nternet.
\newblock \emph{ACM SIGCOMM CCR}, 44\penalty0 (3):\penalty0 58--65, 2014.

\bibitem[Zhang et~al.(2015)Zhang, Chen, Bai, Han, Tian, Wang, Guan, and
  Zhang]{eurosys15}
H.~Zhang, K.~Chen, W.~Bai, D.~Han, C.~Tian, H.~Wang, H.~Guan, and M.~Zhang.
\newblock Guaranteeing deadlines for inter-datacenter transfers.
\newblock In \emph{Proc. EuroSys}, 2015.

\bibitem[Zhang et~al.(2012)Zhang, Xu, Hu, Liu, Guo, and
  Wang]{zhang2012profiling}
X.~Zhang, Y.~Xu, H.~Hu, Y.~Liu, Z.~Guo, and Y.~Wang.
\newblock Profiling skype video calls: Rate control and video quality.
\newblock In \emph{IEEE INFOCOM}, 2012.

\end{thebibliography}
\bibliographystyle{abbrvnat}
}{
}

\end{document}